\documentclass[final, 3p, 12pt, times, authoryear]{elsarticle}

\usepackage{amsmath}
\usepackage{amssymb}
\usepackage{amsfonts}
\usepackage{commath}
\usepackage{microtype}

\usepackage{subcaption}

\usepackage{booktabs}

\newcommand\tens[2][2]{
\ifthenelse{#1=4} 
{\ensuremath{\mathbb{#2}}}
{\ensuremath{\boldsymbol{#2}}}}

\newcommand{\mrm}[1]{\mathrm{#1}}

\renewcommand{\div}[1]{{\rm div }\left( #1 \right)}

\newcommand{\T}{^{\sf T}}

\newcommand*{\defeq}{\mathrel{\vcenter{\baselineskip0.5ex \lineskiplimit0pt
                     \hbox{\scriptsize.}\hbox{\scriptsize.}}}%
                     =}

\newcommand{\lb}{\left(}
\newcommand{\rb}{\right)}

\newcommand{\Secref}[1]{Section~\ref{#1}}  
\DeclareGraphicsExtensions{.pdf}

\journal{International Journal of Solids and Structures}

\usepackage{color}
\usepackage[normalem]{ulem} 

\begin{document}

\begin{frontmatter}



\title{Size and Disorder Effects in Elasticity of Cellular Structures: From Discrete Models to Continuum Representations}

\author[mainadress]{Stefan Liebenstein \corref{mycorrespondingauthor}}
\cortext[mycorrespondingauthor]{Corresponding author}
\ead{stefan.liebenstein@fau.de}
\author[mainadress,TUBAF]{Stefan Sandfeld}
\author[mainadress,Chengdu]{Michael Zaiser}

\address[mainadress]{Institute of Materials Simulation (WW8), Friedrich-Alexander University Erlangen-N\"urnberg (FAU)
, Dr.-Mack-Strasse 77, 90762 F\"urth, Germany}

\address[TUBAF]{Chair of Micromechanical Materials Modelling (MiMM),
	Institute of Mechanics and Fluid Dynamics,
	Technische Universit\"at Bergakademie Freiberg,
	Lampadiusstrasse 4,  09599 Freiberg,
	Germany}

\address[Chengdu]{Department of Engineering and Mechanics, Southwest Jiaotong University,  Chengdu, P.R. China}

\begin{abstract}
Cellular solids usually possess random microstructures that may contain a characteristic length scale, such as the cell size. This gives rise to size dependent mechanical properties where large systems behave differently from small systems. Furthermore, these structures are often irregular, which not only affects the size dependent behavior but also leads to significant property variations among different microstructure realizations. The computational model for cellular microstructures is based on networks of Timoshenko beams. It is a computationally efficient approach allowing to obtain statistically representative averages from computing large numbers of realizations. For detailed analysis of the underlying deformation mechanisms an energetically consistent continuization method was developed which links the forces and displacements of discrete beam networks to equivalent spatially continuous stress and strain fields. This method is not only useful for evaluation and visualization purposes but also allows to perform ensemble averages of, e.g., continuous stress patterns -- an analysis approach which is highly beneficial for comparisons and statistical analysis of microstructures with respect to different degrees of structural disorder.
\end{abstract}


\end{frontmatter}

\section{Introduction}
Cellular solids are a class of lightweight materials which exhibit mass specific mechanical properties superior to their bulk counterparts. Their properties are strongly influenced by geometrical characteristics of the cellular microstructure%
\footnote{In the following, we denote by 'microstructure' internal sub-structures of a material which are characterized by length scales well below the size of a specimen or device, the geometry of which defines the 'macrostructure'.}.
Cellular solids are commonly encountered in complex structures in nature, as for example in cancellous bone, cork or snow. But also man made devices for industrial applications benefit from exploiting microstructure design principles that are analogous to such  naturally occurring materials. Typical examples are (metallic) foams, which may either possess a random microstructure or consist of highly regular, periodic arrangements of unit cells (such as e.g. in honeycomb structures). 
In general, one can distinguish between open- and closed-cellular foams (\cite{gibson1999}): whereas the microstructure of the former consists of a network of connected struts, the latter consists of thin, quasi two-dimensional walls. These porous, network-like microstructures result in low density, high mechanical energy absorption and good weight specific bending stiffness as compared to bulk materials. The details of the microstructure are controlled by the manufacturing process. This can be a natural growth process for biological materials or technological processes such as extrusion, casting, foaming, or additive manufacturing. Often these processes result in microstructures with significant fluctuations in local materials properties and/or geometry, hence microstructural randomness is an important factor influencing the mechanical performance of most solid foams. Furthermore, the characteristic length scales of the cellular microstructures (cell size or bond length) do in general not scale in proportion with the system size, giving rise to mechanical size effects.
First experiments for studying elastic size effects were conducted by \cite{lakes1983, lakes1986}  who investigated quasi-static bending and torsion of beam-like specimens of polymeric foams. The investigation showed that, for comparable density, bigger beams exhibit lower torsion and bending stiffness compared to smaller beams. Other results by \cite{brezny1990} for reticulated vitreous carbon, \cite{choi1990} for cortical bone and \cite{anderson1994} for closed-cell polymethacrylimide foams indicated an opposite behavior where smaller systems were found to be softer than larger ones. 
Similar findings -- smaller systems are elastically weaker -- were reported by \cite{andrews2001} in uniaxial compression of open and closed cell aluminium foams and by \cite{bastawros2000} and \cite{jeon2005} for closed cell aluminium foams. To the knowledge of the authors, stiffness size effects under shear loading were rarely investigated with the exception of \cite{rakow2004,rakow2005} who reported softer behavior when cutting out and shear testing sub-regions of decreasing size from  a larger specimen. However, they stated that this is not a general result, as they also observed small samples which were stiffer -- an observation they attributed to an increasing scatter in materials properties with decreasing sample size.

It should be noted that besides elastic size effects, i.e. size dependency of elastic stiffness, plasticity size effects (size dependency of yield or tensile strength) are a major subject of research (e.g. \cite{andrews1999}, \cite{onck2001}, \cite{kesler2002}, \cite{chen2002}, \cite{jeon2005}). The mechanisms which lead to these size effects are fundamentally different -- unlike yield or failure, elasticity is hardly ever governed by weakest links -- and we we do not investigate them in this work. Another important topic of research is the dependency of mechanical properties on density  (\cite{silva1995}, \cite{gibson1999}, \cite{andrews1999}, \cite{zhu2000}, \cite{roberts2001}). The results are also important with regard to size effects or effects of randomness as one must ensure that, when comparing data from different specimens, systematic differences in density do not influence the results.

Detailed experimental characterization and analysis of the structure-property relations of -- possibly highly disordered -- cellular solids is non-trivial for a number of reasons: First, the geometrical details of the microstructure are usually not a priori known and must be obtained e.g. from 3D computer tomography (see e.g.  \cite{jang2008}), ideally even in-situ under load. Obtaining stresses, strains or energies/energy densities is even more difficult and can usually only be done through post-processing of CT data or by means of digital image correlation of specimen surface images (see e.g. \cite{bart-smith1998}). Another important point is that a statistically meaningful investigation of disordered systems requires a sufficient number of samples, representing different realizations of the disordered microstructure, to be prepared and tested. For highly disordered microstructures this may require investigation of hundreds of samples, which is very time consuming to do in experiments.  

Modelling and simulation approaches, on the other hand, suffer from time restrictions to a much lesser extent and may therefore offer the possibility to analyze in detail the interplay between different microstructure characteristics and the resulting mechanical response. For this purpose, a number of different simulation approaches can be used. The most accurate, but also computationally most expensive approach is to fully resolve the microstructure in all geometrical detail and then to simulate the mechanical response in three dimensions. Often, it is feasible to idealize cellular microstructures as networks of shells or beams. The significant reduction in computational cost for generating and meshing such structures as well as for their simulation makes this approach particularly attractive. This idea to represent a material with a network of beams or as a lattice goes back to \cite{hrennikoff1941}. Because of its simplicity and numerical efficiently the method became popular for studying fracture and elasticity of continua and especially concrete \citep{thorpe1990, roux1985, schlangen1997,herrmann1989}.
In the context of cellular solids it has been applied for example by \cite{onck2001} for the analysis of size effects in honeycomb structures, by \cite{roberts2001} for the study of density dependent elastic properties in 3D closed-cellular solids and by \cite{zhu2000} for investigating the influence of density and microstructural irregularity on the elastic properties of 2D structures. 
For a review of lattice/beam models in micromechanics see \citep{ostoja2002lattice, ostoja2007microstructural}. An efficient simulation setup is particularly important for statistical analysis of random structures and investigation of the associated size effects, where a large number of simulations with different realizations of a disordered microstructure may need to be performed. 

An alternative to fully resolved models of the material microstructure consists in using approaches where the microstructure is averaged out and replaced by a continuous effective medium. Continuum models which contain an instrinsic length scale and therefore can represent size effects are for example so-called micromorphic continua developed by Eringen and coworkers ( \cite{eringen1964},  \cite{eringen1966}, \cite{eringen1999}) which were used for example by \cite{dillard2006}, \cite{diebels2002} \cite{tekoglu2005} in the context of metallic foams. 
The major benefit of continuum models is the reduced computational cost which allows to simulate complex geometries and loading conditions. Drawbacks consist in reduced accuracy, but even more in the requirement to formulate appropriate constitutive laws, to identify the corresponding parameters which for fully micromorphic theories may be numerous, and to define appropriate boundary conditions for the higher-order stress and strain variables.

In this paper we study the size-dependent elastic behavior of cellular structures by fully resolved simulations. Influences of boundary conditions and microstructural randomness are investigated and statistically analyzed. For detailed investigation of local deformation patterns we introduce a new, energetically consistent continuization scheme which allows us to represent the deformation state of a discrete network-like microstructure of beams in terms of spatially continuous stress and strain fields. We start in \Secref{sec:model} by introducing the rules we use for generating the discrete microstructures, and defining the corresponding micro-scale material properties as well as the boundary conditions in our simulated deformation tests. In \Secref{sec:homogenization} we present the derivation of the continuization scheme and its verification. Results for macroscopic system response as well as micro-scale deformation patterns are shown in \Secref{sec:results}. Finally we discuss the results in \Secref{sec:discussion} and link some aspects of the system-scale deformation behavior to statistical features of the underlying micro-scale deformation patterns before we conclude this work in \Secref{sec:conclusion}.

\section{Model description}
\label{sec:model}
We consider two-dimensional systems and envisage the cellular microstructures as network-like arrangements of struts. The aspect ratio between strut length $l$ and thickness $w$ is assumed sufficiently large (i.e. $l/w \geq \approx 5$) such that we may approximate the microstructure as a network of interconnected, quasi-one-dimensional beams. Creating such networks  is a well-defined and computationally inexpensive process. In FEM simulations the beams can be described by one-dimensional elements of which only the positions and connectivity of the end points need to be specified. This results in a very significant reduction in degrees of freedom as opposed to solid two- or three-dimensional elements.
We note that in real materials struts often show a thickening towards junctions. To take this into account and improve the model it would be possible to use beam elements with varying cross-section as presented e.g. by \cite{auricchio2015}. Within the present work, however, we consider beam elements of constant cross-section which can be envisaged as Timoshenko beams. This implies that we make the following simplifying assumptions (see e.g. \cite{zienkiewicz2005b}): (i) each beam has constant cross section; (ii) the dimensions of the cross-section are small compared to the beam length; (iii) the cross-sections remain planar and their shape and size does not change under loading, but they do not necessarily stay perpendicular to the beam axis. Displacements along the beam axis are approximated by shape functions with a linear Ansatz. To prevent shear locking a cubic Ansatz is chosen for beam deflection and cross-section rotation. An overview of the governing equations of the Timoshenko beam can be found in \ref{appendix:eq_timoshenko_beam}.

\subsection{Microstructure generation}
The microstructure of real open-cellular foams is often not regular and exhibits random variations which result from the manufacturing process. One of the most common methods to computationally generate random cellular geometries is the Voronoi tessellation (\cite{zhu2000, gibson1999}).
As discussed by \citet{boots1982, vanderburg1997} it reasonably represents a foaming process under the assumptions that
\begin{enumerate}
\item all nuclei appear simultaneously,
\item the nuclei remain fixed in position throughout the growth process,
\item each nucleus grows isotropically, i.e., at the same rate in all directions,
\item the growth stops for each cell whenever it comes into contact with a neighboring cell.
\end{enumerate}
 Although the tessellation as well as the whole procedure for the microstructure generation is easily generalized to any dimension we focus here on the 2D case. In a first step spatially distributed seed points are generated in the investigated domain. Then to each seed ${\cal S}$ a cell is assigned which we define as the set of all points whose Euclidean distance to ${\cal S}$ is smaller than that to any other seed. These so-called Voronoi cells are polygons (in 3D: convex polyhedra) which tessellate the plane. The 2D Voronoi tessellation can then be used to generate a cellular microstructure by placing beams on the cell boundaries. The distribution of the Voronoi seeds is a determining factor for the resulting microstructure. As two limit cases we have regular seeding (seeds are placed on a periodic lattice), and a completely random distribution of seed points according to a Poisson point process. A typical 2D example for structures obtained from regular seeding are hexagonal grids (honeycombs) which result from seeds being placed on a triangular lattice. 

The emphasis of this work is on systems with tunable degree of randomness. \cite{zhu2000} proposed a method for creating 2D tesselations, which was used for example by \cite{onck2001} and \cite{tekoglu2005}. They start with an initial random seed, followed by generating a sequence of new random seeds, each of which is only accepted if it has a minimal distance $\delta_{\mrm{min}}$ to the initial seeds. The minimal distance acts as an order parameter which controls the randomness of the system. For $\delta_{\mrm{min}}=0$ the seeds are distributed independently, corresponding to a Poisson random process. The other limit is a densest packing of identical cells. For 2D this corresponds to a tessellation with honeycombs of equal edge length (\cite{hales2001}), in which case $\delta_{\mrm{min}}$ corresponds to the distance $\Delta p$ between the sites of a triangular lattice. However, the numerical generation of highly ordered systems ($\delta_{\mrm{min}} >0.8 \Delta p$) from a random seeding procedure is exceedingly unlikely for simple reasons of (information) entropy. Therefore we use here a different approach (see e.g. \cite{vanderburg1997,garboczi1987}) for generating Voronoi tessellations with tunable degree of randomness.

\begin{figure}
\subcaptionbox{$\beta= 0$\label{fig:honeycomb_microstructure}}{\raisebox{0.005\linewidth}
{\includegraphics[height=.271\linewidth]{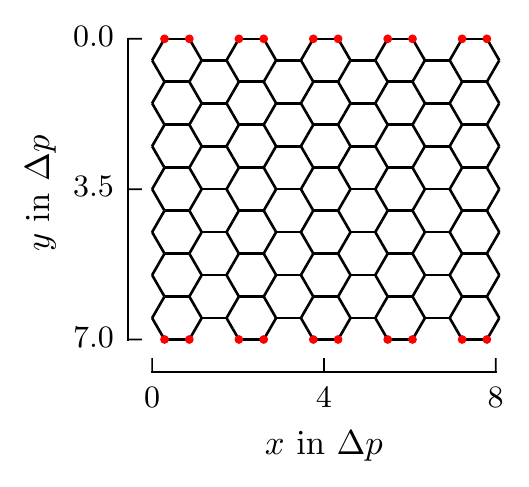}}}
\subcaptionbox{$\beta= 0.1$\label{fig:microstructure0.1}}
{\includegraphics[height=.27\linewidth]{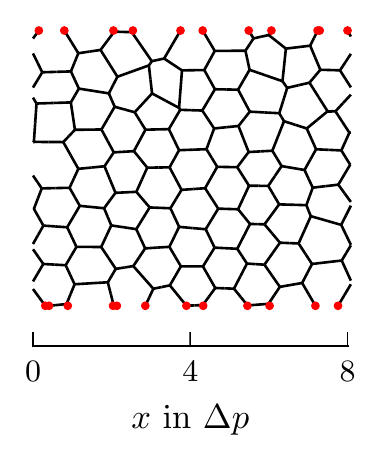}}
\subcaptionbox{$\beta= 0.3$}
{\includegraphics[height=.27\linewidth]{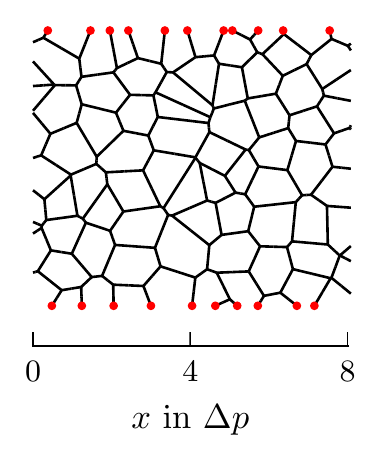}}
\subcaptionbox{$\beta= 5.0$}
{\includegraphics[height=.27\linewidth]{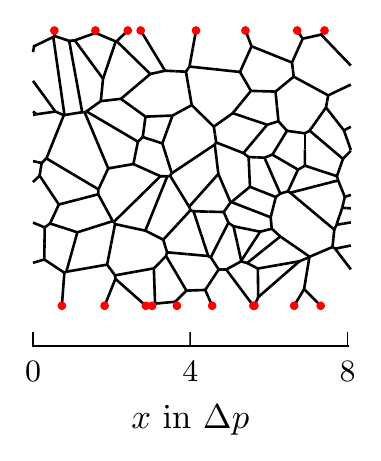}}
\caption{Microstructures for different parameters $\beta$, system height $H=7\Delta p$. The nodes at which boundary conditions are applied are indicated by red dots.}\label{fig:microstructures}
\end{figure}

In this approach seeds are first located on a triangular lattice with lattice constant $\Delta p$. The position $\tens{p}$ of each seed point is then perturbed by a stochastic variable $\delta \tens{p}$. Therefore the full range between fully regular and completely random systems can be generated with the same computational effort by simply varying the statistical rules for choosing $\delta \tens{p}$. For the direction of the perturbation an isotropic orientation distribution is used, whereas the distribution of perturbation distances $|\delta \tens{p}|$ is assumed exponential, 
\begin{align}
f\lb \frac{|\delta \tens{p}|}{\Delta p},\beta \rb = \frac{1}{\beta} \exp \lb -\frac{1}{\beta}\frac{|\delta \tens{p}|}{\Delta p} \rb,
\end{align}
where the disorder parameter $\beta>0$ defines both the mean value and the standard deviation of the distribution. For $\beta \gg 1$ the final position of the seeds is exclusively governed by the stochastic process, which implies that the seeds are again distributed according to a Poisson point process. The influence of the parameter $\beta$ on the microstructure can be seen in \figref{fig:microstructures}.

In our deformation simulations we consider specimens of rectangular shape. The triangular lattice used in the first step of our seeding procedure is oriented in such a manner that one side of its unit cell aligns with the $y$ axis, which in graphical representations runs across a vertical specimen boundary. The resulting Voronoi tessellation is the dual of the seeding and an arrangement of regular honeycombs, which cell sides are aligned with the $x$ axis (cf. \figref{fig:honeycomb_microstructure}).
 To compare systems with different $\beta$ it must be ensured that the areal density of seeds is maintained. This might be relevant in situations where the perturbation results in seeds being displaced outside the investigated domain, leading to a reduction in seed density near the specimen boundaries. Therefore we use periodic boundary conditions to map such seeds back into the specimen domain. After seeding we determine the associated Voronoi cells and place beams on the cell boundaries. If a cell boundary intersects the sample boundary, the beam is terminated at the intersection point. 

\subsection{Microstructural material properties}
\label{sec:microproperties}
The mechanical response of the systems depends on the morphology of the microstructure but also on the properties of the beams. In the following we assume that all $N_\mrm{B}$ beams of varying lengths $l_i$ have the same in-plane width $w$ and out of plane thickness $t$. Thus the total volume covered by the beams is 
\begin{align}
V_{\mrm B} = t w \sum_{i=1}^{N_\mrm{B}} l_i.
\end{align}
Macroscopically the system has the same thickness $t$ as the beams, its width is denoted by $W$  and its height by $H$. Thus the relative density is given by
\begin{align}
\label{eq:density_scaling}
\rho_{\mrm{rel}} = \frac{V_{\mrm B}}{ t W H} = \frac{w}{W H} \sum_{i=1}^{N_\mrm{B}} l_i.
\end{align}

Since we do not study density dependent behaviour, the relative density must be the same for all systems. We ensure a relative density of $\rho_{\mrm{rel}} = 0.1$ by propotional scaling of $W$ and $H$. The beams are assumed to have linear elastic material properties with Young's modulus $E_{\mrm{B}} = 0.1 \text{ GPa}$ and Poisson's ratio $\nu_{\mrm{B}}= 0.3$. The cross-section is assumed quadratic with $t = w = 0.05 \Delta p$ and a beam shear coefficient as proposed by \cite{cowper1966} of $\kappa = (10+ 10\nu_{\mrm{B}} ) /(12+11 \nu_{\mrm{B}}) \approx 0.85$ (cf. Appendix A).

\subsection{Macroscopic boundary conditions}
\label{sec:BC}
Our investigation considers two distinct loading conditions --~uniaxial compression and simple shear~-- which are imposed in a displacement driven manner. In accordance to experiments (\cite{andrews2001}, \cite{tekoglu2011}) frictionless grip for compressive loading and stick grip for simple shear loading is chosen. Under compressive loading the beams are not clamped at the boundaries, allowing for free rotation (characterized by the rotation vector $\tens{\phi} = \phi \tens{e}_z$), cf. \figref{fig:beam_kinematics}. The simple shear loading is defined by no beam rotations at the top and bottom boundaries and free rotations at the sides.
Prescribed nodal displacements $\bar{u}$ are applied to all nodes on the top surface of the discrete beam system. $\bar{u}$ is chosen such that an isotropic continuum system would have an engineering strain of $| \varepsilon^* |=0.05$.
This imposed strain is on the upper limit or outside of the elastic regime for real materials, especially if we consider that local strains might be even higher. However, the chosen elastic model scales linearly and thus is representative for any proportional elastic loading, i.e., values for lower strains can be obtained by simple multiplication with the strain ratio.
An overview of the applied boundary conditions is given in Table~\ref{tb:Dirichlet BC}.

\begin{table}[tbp]
\centering
\begin{tabular}{ l l l l l l}
\toprule
 Loading & Bottom ($y=0$) & Top ($y=H$) & Left ($x=0$) & Right ($x=w$) \\
  \midrule
  Compression & $\bar{u}_y=0$ & $\bar{u}_y= H / 0.05$ & free & free \\
  Simple Shear  & $ \bar{u}_x=0, \phi=0$  & $ \bar{u}_x= H/ 0.05 ,\phi=0$ & free& free\\
  \bottomrule
\end{tabular}
  \caption{Boundary conditions for the two  investigated loading conditions.}
  \label{tb:Dirichlet BC}
\end{table}

Studying size effects requires comparison of systems which are geometrically similar, i.e have the same width to height ratio. 
Furthermore it is necessary that the relative density of the system remains the same (cf. \Secref{sec:microproperties} ). 
A scaling which allows for comparison with closed honeycombs (\figref{fig:honeycomb_microstructure}) is given by the height and width values
\begin{align}
\label{eq:scaling1}
H(i) &=  \Delta p (4  + 3 i)\qquad \text{and}\qquad
W(j)= \frac{ 2}{\sqrt{3}} \Delta p(4 + 3 j),
\end{align}
with $i,j \in \mathbb{N}$. Throughout the following investigation we shall assume, unless otherwise stated, that $i=j$, corresponding to a ratio $W/H = 2/\sqrt{3} =: R_0$. 

Depending on the way boundaries are located, regular honeycombs may differ very substantially from systems with a small degree of disorder. In a regular honeycomb, the position of beam endpoints may exactly coincide with the system boundaries such that all boundary cells are closed (\figref{fig:honeycomb_microstructure}). In slightly perturbed structures, on the other hand, depending on the positions of the seeds at the boundary the corresponding cells may either be closed or (partially) open (\figref{fig:microstructure0.1}). This could lead to the unphysical result that even an infinitesimal perturbation of a regular honeycomb might produce a significant reduction in stiffness. To mitigate this problem we choose the position of the boundaries with respect to the seeds at random. Our regular reference microstructure has a periodicity of $\Delta p$ in $y$-direction and of $(1+\frac{ 2}{\sqrt{3}} )\Delta p$ in $x$-direction. To define our system boundaries we chose random cut-outs where the system boundaries are shifted with respect to the underlying seed lattice by random amounts
\begin{align}
\Delta y = 0.5 \Delta p  \,R_{\mrm h}(-1,1) \qquad \text{and} \qquad \Delta x = (0.5+\frac{ 1}{\sqrt{3}} )\Delta p \, R_{\mrm w}(-1,1),
\end{align}
where $R_{\mrm w}(-1,1), R_{\mrm h}(-1,1)$ are random numbers which are uniformly distributed between $-1$ and $1$.  It is important to note that different cut-out positions may lead to slightly different total beam lengths.
This in turn causes us to re-scale the system size such that a relative density of $\rho_\mrm{rel}=10\%$ is always preserved.

\subsection{Macroscopic system response}
\label{sec:sys_response}
For comparing the macroscopic mechanical response of different systems we investigate global stiffness parameters of the system. 
To distinguish between macroscopic properties (e.g. the average engineering strain of the system) and microstructure properties (e.g. axial beam strains) all macroscopic properties are in the following denoted by the superscript ${}^*$. The averaged, macroscopic displacements at the top and bottom specimen boundaries are given by
\begin{align}
\tens{u}^{* \mrm T} = \tens{u}^*(y=H) = \frac{1}{N_{\mrm T}} \sum_{\alpha=1}^{N_{\mrm T}} \tens{u}^\alpha (y=H),\\
\tens{u}^{* \mrm B} = \tens{u}^*(y=0) = \frac{1}{N_{\mrm B}} \sum_{\alpha=1}^{N_{\mrm B}} \tens{u}^\alpha (y=0),
\end{align}
where $N_{\mrm T}$ and $N_{\mrm B}$ are the numbers of surface nodes of the beam network at the top and bottom boundaries, respectively,
and $\tens{u}^\alpha(y)$ denotes the displacement vector of surface node number $\alpha$ at position $y$ (cf. \figref{fig:microstructures}). The macroscopic axial and shear strains are defined as
\begin{align}
\varepsilon_{yy}^* = \frac{|u_y^{* \mrm T}- u_y^{* \mrm B}|}{H} \qquad \text{and}\qquad
\varepsilon_{xy}^* = \frac{|u_x^{* \mrm T}- u_x^{* \mrm B}|}{H}.
\end{align}
The average stiffnesses $C^*_\mrm{c,s}$ for uniaxial compression and simple shear are evaluated by calculating the sums of all load-induced reaction forces $R^\alpha_x$ and $R^\alpha_y$ in $x$ and $y$ direction, and dividing them by the loaded (macroscopic) area $A_{\mrm S} = W t$ and the respective averaged strain components $\varepsilon^*_{yy}, \varepsilon^*_{xy}$:
\begin{align}
\label{eq:macro_stiffness}
C^*_{\mrm c} &= \frac{1}{A_{\mrm S} \varepsilon^*_{yy}} \sum_{\alpha=1}^{N_\mrm{R}} R^\alpha_{y}  \qquad \text{for uniaxial compression} \\
C^*_{\mrm s} &= \frac{1}{A_{\mrm S} \varepsilon^*_{xy}} \sum_{\alpha=1}^{N_\mrm{R}} R^\alpha_{x} \qquad \text{for simple shear}.
\end{align}
Note that the stiffnesses $C^*_\mrm{c,s}$ are not intrinsic material properties as they depend on the macroscopic geometry of the sample (cf. \Secref{sec:glob_results}).

\section{Energetically consistent discrete-to-continuum transition}
\label{sec:homogenization}
The mechanical response of our model systems arises from deformation of the beam network representing the microstructure. In this network, forces and displacements are evaluated only at the nodes of the beam elements. In different realizations of the same random microstructure these nodes are located at different points in space. If we want to average over different realizations of the same random microstructure, to make comparisons between different microstructures, or to analyze sample-scale deformation patterns, we need to perform a transition from displacements and forces located at discrete nodal points to a representation of the deformation state in terms of spatially continuous stress and strain fields. The same is true if we want to compare results from our discrete microstructure model to those derived from (generalized) continuum frameworks.   

\begin{figure}
\centering
\includegraphics[width=0.3\textwidth]{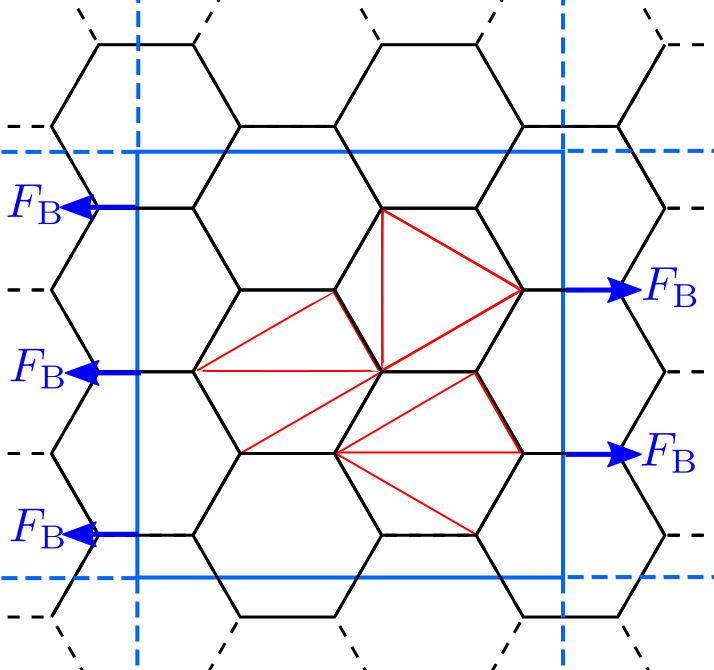}
\caption{Section from the middle of a regular honeycomb system. In blue a microstructure independent control volume which results in artificial stress oscillations. In red: possible triangulations as proposed by \cite{tekoglu2011}.}
\label{fig:hom_schemes}
\end{figure}

A continuum representation of the local mechanical state of the system should be based on the available data (nodal forces and displacements) and should "guess" as little information as possible by interpolation. To this end the system domain is decomposed into smaller sub-domains in which discrete and continuous formulations are matched locally. We require that these so-called control volumes are (i) in force and moment equilibrium, (ii) can be defined for regular and random microstructures, and (iii) allow for a formulation of the deformation state in terms of stress and strain that is energetically consistent with the discrete system. 
Schemes which use fixed rectangular subdomains and then compute stresses from the interface forces are microstructure independent but will in general produce artificial oscillations of the stress fields on the scale of the control volumes. A simple example can be seen in \figref{fig:hom_schemes} where a section from a regular honeycomb system is shown with a homogenization subdomain. Assume that the system is homogeneously compressed in the horizontal direction as shown in the figure. It can be seen that force equilibrium of the given control volume of size $t \times L_x \times L_y$ requires the stress gradient $\Delta \sigma_{xx}/\Delta x = - F_B/(t L_x L_y)$ in $x$ direction to be balanced by an equal but opposing shear stress gradient $\Delta \sigma_{xy}/\Delta y = F_B/(t L_x L_y)$ in the perpendicular direction. This stress gradient oscillates: it will occur with the same magnitude but opposite sign in the adjacent control volumes. Since the stress state is actually homogeneous -- all cells of the honeycomb structure are loaded in an identical manner -- such oscillating stresses must be envisaged as artifacts of the microstructure-independent control volume. 
A cell-wise procedure, e.g. by dividing each Voronoi cell individually into triangles as proposed by \cite{tekoglu2008, tekoglu2011} and shown in \figref{fig:hom_schemes} works well for strain mappings, but is not suited for stress continuization since it is not clear how to evaluate without ambiguity force couples acting on the control volumes from the forces acting on the junction points.  

To overcome these limitations we propose a novel approach which allows to compute stresses and strains in an energetically consistent manner while using control volumes that reflect the local microstructure morphology and thus minimize artefacts. Furthermore, this method is suitable for regular as well as for random microstructures. The idea is inspired by a very similar approach commonly used with discrete element methods (see for example \cite{christoffersen1981}, \cite{mehrabadi1982}, \cite{bagi1996}). The main idea is that energy consistency during transition from a discrete network to an equivalent continuum representation is guaranteed by using the principle of virtual work: For any given control volume $V_\mrm{c}$, the sum of the internal virtual work $\delta W_\mrm{b}^\mrm{int}$ of all $N_\mrm{b}$ beams therein must be the same as the internal virtual work $\delta W_\mrm{c}^\mrm{int}$ done by the continuum,  
\begin{align}
\delta W_\mrm{c}^\mrm{int} =\sum_{\alpha=1}^{N_\mrm{b}} \delta W_\mrm{b}^{\mrm{int}, \alpha} \quad \text{in } V_\mrm{c}.
\end{align}
As the systems are in static equilibrium the external virtual work is equal to the internal virtual work so that
\begin{align}
\label{eq:beam_continuum_balance}
\delta W_\mrm{c}^\mrm{int} =\sum_{\alpha=1}^{N_\mrm{b}} \delta W_\mrm{b}^{\mrm{ext}, \alpha} \quad \text{in } V_\mrm{c}.
\end{align}

\subsection{Energy of the Timoshenko Beam}
\label{sec:beamenergy}
\begin{figure}[htp] \centering
	\includegraphics[width=0.95\textwidth]{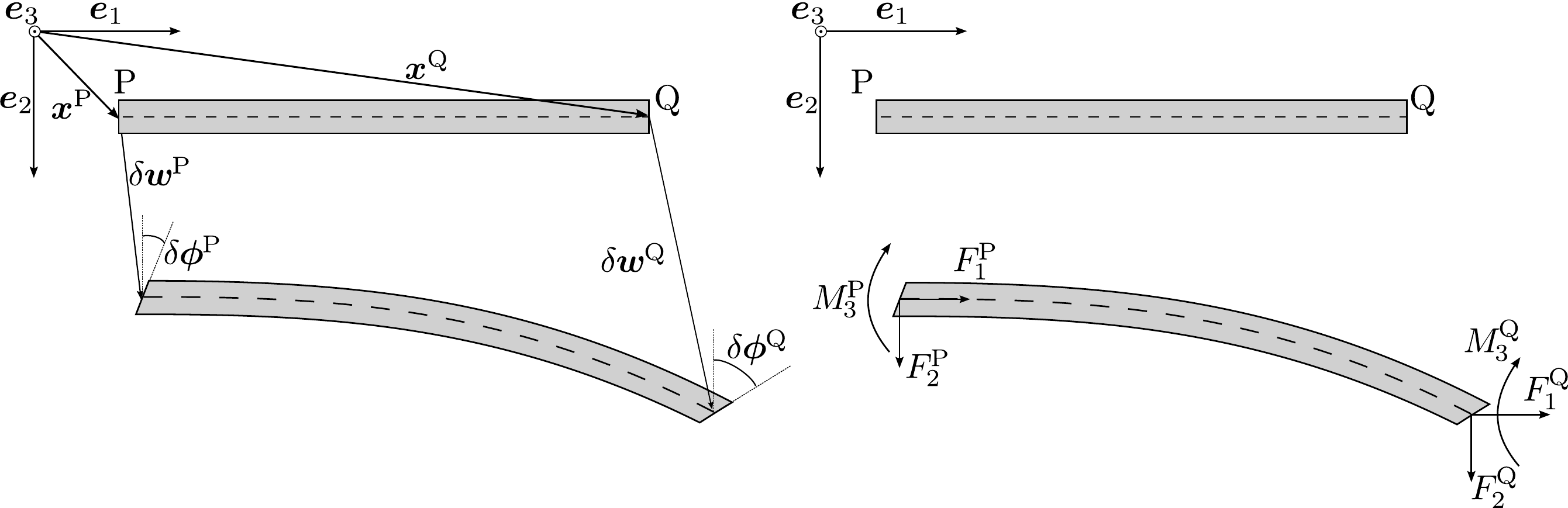}
	\caption{Kinematics (left) and forces (right) in 2D for the Timoshenko beam. For better visualization the rotations are largely exaggerated.}
	\label{fig:beam_kinematics}
\end{figure}

For derivation of the continuum stresses we first consider a single discretized Timoshenko beam in 2D as used in our finite element simulation, cf. \figref{fig:beam_kinematics} and \ref{appendix:eq_timoshenko_beam}. 
The beam is defined by the end points $P$ and $Q$ with the corresponding beam forces $\tens{F}$ and torques $\tens{M}$. The external virtual work $\delta W_\mrm{b}^\mrm{ext}$ done by the virtual displacements $\delta \tens{w}$ and the virtual rotations $\delta \tens{\phi}$ is given by
\begin{align}
\label{eq:timoshenko_beam}
\delta W_\mrm{b}^{\mrm{ext}} = \tens{F}^\mrm{P} \cdot \delta \tens{w}^\mrm{P} + \tens{F}^\mrm{Q}\cdot \delta \tens{w}^\mrm{Q} + \tens{M}^\mrm{P}\cdot \delta \tens{\phi}^\mrm{P}+ \tens{M}^\mrm{Q}\cdot \delta \tens{\phi}^\mrm{Q},
\end{align}
where $\cdot$ denotes a single contraction between two vectors or tensors.
The so-called beam vector connects $P$ and $Q$:
\begin{align}
\label{eq:beamvec}
\tens{l} = \tens{x}^\mrm{P}-\tens{x}^\mrm{Q}.
\end{align}
Furthermore we define the virtual separation $\delta \tens{\Delta}^w$ as the difference of the virtual displacements
\begin{align}
\label{eq:rel_disp}
\delta \tens{\Delta}^w = \delta \tens{w}^\mrm{P} - \delta \tens{w}^\mrm{Q}.
\end{align}
Because the beam is in static equilibrium it follows with the cross product $\times$ for forces and moments:
\begin{align}
\label{eq:beamforces}
\tens{F} &  \defeq \tens{F}^\mrm{P} = -\tens{F}^\mrm{Q},\\
\tens{M} & \defeq \tens{M}^\mrm{P} = - \tens{M}^\mrm{Q} - \tens{l} \times \tens{F}^\mrm{Q} = - \tens{M}^\mrm{Q}+\tens{l} \times \tens{F}.
\end{align}
Therefore we obtain the expression for the virtual work from \eqref{eq:timoshenko_beam} as
\begin{align}
\begin{split}
\delta W_\mrm{b}^\mrm{ext} &= \tens{F}\cdot \delta \tens{\Delta}^w + \tens{M}\cdot \delta \tens{\phi}^\mrm{P} + \lb \tens{l} \times \tens{F} - \tens{M}  \rb\cdot \delta \tens{\phi}^\mrm{Q}.
\end{split}
\end{align}

\subsection{Derivation of energy equivalent stress fields}

\begin{figure}[htbp]
	\centering
	\includegraphics[width=0.5\textwidth]{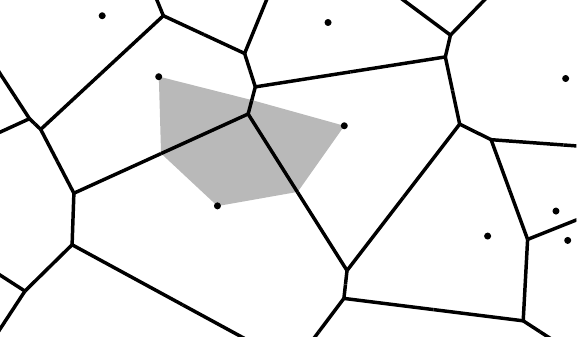}
	\caption{Two dimensional microstructure with center points of the Voronoi tessellation a control volumes (shown in grey).}
	\label{fig:tessellation}
\end{figure}

The energy equivalence \eqref{eq:beam_continuum_balance} must hold for all local control volumes. To allow for a meaningful computation of forces in conjunction with ensuring equilibrium of forces and moments the boundaries of these control volumes must be chosen such that they intersect with at least three beams that are connected to the same node of the beam network. For the construction of polygonal control volumes we propose the following method:
\begin{enumerate}
\label{poly_construction}
\item Pick an arbitrary junction point which has three (or more) connecting beams as base.
\item Choose the midpoints between the base point and the connected junctions  as corners of the polygons. This assures that all beams are covered by the polygons. 
\item Choose the center points of the Voronoi tessellation as additional corner points to ensure a tessellation of the whole domain.
\end{enumerate} 
This procedure is carried out for all junction points leading to full coverage of the system by a tessellation of irregular polygons (compare \figref{fig:tessellation}).

When deriving energy consistent stress variables for these control volumes, an additional difficulty arises from the fact that the beams - and thus the control volumes - may not only deform but also rotate, which may change the elastic energy. Such behaviour can not be captured in classical continuum formulations because material points do not have rotational degrees of freedom. The simplest extension of classical continuum elasticity which allows for an energy consistent formulation by accounting for rotation effects are the so-called Cosserat or micropolar continua (see for example \cite{eringen1999} which we choose as our framework. Introducing the Levi-Civita tensor
\begin{align}
\tens{\epsilon} = \epsilon_{ijk} \tens{e}_i \otimes \tens{e}_j \otimes \tens{e}_k = 
\begin{cases}
+1 & \text{if } (i,j,k) \text{ is } (1,2,3), (2,3,1) \text{ or } (3,1,2), \\
-1 & \text{if } (i,j,k) \text{ is } (3,2,1), (1,3,2) \text{ or } (2,1,3), \\
\;\;\,0 & \text{if }i=j \text{ or } j=k \text{ or } k=i
\end{cases} 
\end{align}
 and denoting the double contraction of two tensors by $:$, the static balance equations for a Cosserat continuum without body forces reads
\begin{align}
\label{eq:linear_momentum}
\div{\tens{\sigma}+\tens{S}} = \tens{0},\\
\label{eq:angular_momentum}
\div{\tens{\mu}} - \tens{S}: \tens{\epsilon}= \tens{0}.
\end{align}
The additional couple stress $\tens{\mu}$ balances the angular momentum introduced by the  skew-symmetric stress tensor $\tens{S} = -\tens{S} \T$, whereas the stress tensor $\tens{\sigma}$ is symmetric. As a consequence the total stress $\tens{\sigma}+\tens{S}$, which corresponds to the Cauchy stress is, unlike in classical continuum theories, no longer symmetric.
Neumann boundary conditions now depend on the traction vector $\tens{t}$ and the vectorial higher order couple traction $ \tens{m}$ acting on the surface:
\begin{align}
\label{eq:traction_bc}
(\tens{\sigma}+\tens{S})\T \cdot \tens{n} &= \tens{t}, \\
\tens{\mu}\T  \cdot \tens{n} &= \tens{m},
\end{align}
where $\tens{n}$ is the normal of the surface $S_{\mrm c}$ of the control volume $V_{\mrm c}$.
With the (continuum) displacements $\tens{u}$ and micro rotations $\tens{\omega}$ the corresponding work conjugated strain measures are
\begin{align}
\tens{\varepsilon} & = \mathrm{sym} \lb\nabla \tens{u}\rb = \frac{1}{2} ( \nabla \tens{u} + \nabla \tens{u} \T), \\
\tens{\varepsilon}^{\mrm R} &= \mathrm{skw} \lb\nabla \tens{u}\rb + \tens{\epsilon}\cdot\tens{\omega}= \frac{1}{2} ( \nabla \tens{u} - \nabla \tens{u} \T)+ \tens{\epsilon}\cdot\tens{\omega}.
\end{align}
The virtual work of the continuum system follows as
\begin{align}
\delta W_{\mrm c} &= \int_{S_{\mrm c}} \tens{t}\cdot \delta \tens{u} \dif A + \int_{S_{\mrm c}} \tens{m}\cdot \delta \tens{\omega} \dif A - \int_{V_{\mrm c}} \lb \tens{\sigma} : \delta \tens{\varepsilon} + \tens{S} : \delta\tens{\varepsilon}^{\mrm R} +  \tens{\mu}: \nabla  \delta \tens{\omega} \rb \dif V=0,
\end{align}
which defines the internal virtual work as
\begin{align}
\delta W_{\mrm c}^\mrm{int} &= \int_{V_{\mrm c}} \lb\lb \tens{\sigma} + \tens{S} \rb :  \nabla  \delta \tens{u} - \tens{S}: \lb\tens{\epsilon}\cdot\delta \tens{\omega} \rb +  \tens{\mu}: \nabla  \delta \tens{\omega} \rb \dif V.
\label{eq:virt_work}
\end{align}
The position and shape of the control volumes is defined by the positions of the beams. A change of the beam end positions $Q^\alpha$ results in a strain and a rotation around the junction point $\tens{x}^\mrm{P}$. For regular systems the centroid of the control volume  $\tens{x}^\mrm{c} = \frac{1}{V_{\mrm c}}\int_{V_{\mrm c}} \tens{x} \dif V$  coincides with this junction point. Irregularity, which is governed by the perturbation $\delta \tens{p}$, leads to a mismatch between the two points. As the continuous system rotates around $\tens{x}^\mrm{c}$ in irregular control volumes the center of rotation is not the same for the continuous and the discrete system. To circumvent this problem we notice, that the two points are often very close to each other, so that we
 approximate the centroid by
\begin{align}
\tens{x}^\mrm{c} \approx \tens{x}^\mrm{P},
\label{eq:centroid_pos}
\end{align}
which can be envisaged as a zeroth-order expansion of the centroid position with respect to $\delta \tens{p}$. The truncated expansion is exact for regular systems and ensures that (i) the continuous and the discrete system have the same center of rotation and (ii) a pure rigid-body motion of the system, i.e. 
\begin{align}
\mathrm{sym}\lb {\nabla \tens{u}}\rb = \tens{0},\qquad \mathrm{skw}\lb {\nabla \tens{u}}\rb = \tens{\epsilon}\cdot\tens{\omega}
\end{align} 
does not exhibit any energy.

We assume that stresses as well as couple stresses are constant in each polygonal control volume. This implies that the corresponding virtual displacement and rotation fields can be written as linear mappings 
\begin{align}
\label{eq:linear_ansatz}
\delta \tens{u} &= \tens{\Psi} \cdot \tens{x} + \tens{q},\\\
\label{eq:rotation_ansatz}
\delta \tens{\omega} & = \tens{\Phi}\cdot \tens{x} + \tens{r},
\end{align}
with the constant mapping tensors $\tens{\Psi}$ and $\tens{\Phi}$ and the constant vectors $\tens{q}$ and $\tens{r}$.
Note that the constant stress/strain assumption is made for the stress/strain field of each control volume separately. The global stress distribution is therefore piecewise constant.
The virtual displacement fields for the junction nodes are then given in terms of the respective position vectors $\tens{x}^\mrm{P}$ and $\tens{x}^\mrm{Q}$ by
\begin{align}
\label{eq:P_disp}
\delta \tens{w}^\mrm{P} = \tens{\Psi} \cdot \tens{x}^\mrm{P} + \tens{q}, 
\quad&\delta \tens{w}^\mrm{Q} = \tens{\Psi} \cdot \tens{x}^\mrm{Q} + \tens{q},\\
\label{eq:P_rot}
\delta \tens{\phi}^\mrm{P} = \tens{\Phi}\cdot \tens{x}^\mrm{P} + \tens{r},
\quad&\delta \tens{\phi}^\mrm{Q} = \tens{\Phi}\cdot \tens{x}^\mrm{Q} + \tens{r}.
\end{align}
Because the virtual micro rotations and the virtual displacements can be considered separately we start by deriving the expressions for the virtual displacements,
\begin{align}
\label{eq:virtual_force_equilibrium}
 \int_{V_{\mrm c}} \lb \tens{\sigma} + \tens{S} \rb : \nabla \delta \tens{u} \dif V = \sum_{\alpha=1}^{N_\mrm{b}}  \tens{F}^\alpha\cdot \delta \tens{\Delta}^\alpha.
\end{align}
With \eqref{eq:rel_disp}, \eqref{eq:linear_ansatz} and \eqref{eq:P_disp} we can write \eqref{eq:virtual_force_equilibrium} as 
\begin{align}
\int_{V_{\mrm c}} \lb \tens{\sigma} +\tens{S} \rb: \tens{\Psi} \dif V = \sum_{\alpha=1}^{N_\mrm{b}} \tens{F}^\alpha\cdot \lb \tens{\Psi}\cdot\tens{l}^\alpha \rb.
\end{align}
Since $\tens{\Psi}$ is arbitrary and constant we obtain
\begin{align}
\int_{V_{\mrm c}}\lb \tens{\sigma} +\tens{S} \rb \dif V = \sum_{\alpha=1}^{N_\mrm{b}} \tens{F}^\alpha \otimes  \tens{l}^\alpha 
\label{eq:stress_force_relation}
\end{align}
for the stress in the volume. From this the average stress in the control volume evaluates as
\begin{align}
\langle \tens{\sigma} +\tens{S}  \rangle_\mrm{c} = \frac{1}{V_{\mrm c}} \int_{V_{\mrm c}} \lb \tens{\sigma} +\tens{S} \rb \dif V, 
\end{align}
and we can write \eqref{eq:stress_force_relation} as
\begin{align}
\langle  \tens{\sigma} +\tens{S}  \rangle_\mrm{c} = \frac{1}{V_{\mrm c}} \sum_{\alpha=1}^{N_\mrm{b}}\tens{F}^\alpha \otimes  \tens{l}^\alpha.
\end{align}
Because $\tens{\sigma}$ is symmetric and $\tens{S}$ skew-symmetric their averages can be considered separately, such that
\begin{align}
\langle \tens{\sigma} \rangle_\mrm{c} &= \mathrm{sym} \lb\frac{1}{V_{\mrm c}} \sum_{\alpha=1}^{N_\mrm{b}}\tens{F}^\alpha \otimes  \tens{l}^\alpha\rb, &
\langle \tens{S} \rangle_\mrm{c} &= \mathrm{skw} \lb\frac{1}{V_{\mrm c}} \sum_{\alpha=1}^{N_\mrm{b}}\tens{F}^\alpha \otimes  \tens{l}^\alpha\rb.
\end{align}
The average of a quantity $(\cdot)$ for the whole system can then be approximated by
\begin{align}
\langle \cdot \rangle_\mrm{S} \approx  \frac{1}{V_{\mrm S}} \sum_{k=1}^{N_{\mrm c}} V_{\mrm c}^k \langle \cdot \rangle_\mrm{c}^k,
\end{align}
where $V_{\mrm S} = \sum_{k=1}^{N_{\mrm v}} V_c^k$ is the system volume and $ N_{\mrm c}$ the number of control volumes in the system. Thus, the system-averaged stress can be approximated by
\begin{align}
\langle \tens{\sigma} +\tens{S} \rangle_\mrm{S} =  \frac{1}{V_{\mrm S}} \sum_{k=1}^{N_{\mrm c}} V_{\mrm c}^k \langle\tens{\sigma} +\tens{S}\rangle_\mrm{c}^k.
\end{align}
Secondly the virtual work equivalence for the virtual rotations is investigated
\begin{align}
\label{eq:virt_rot}
\sum_{\alpha=1}^{N_\mathrm{b}}   \tens{M}^\alpha \cdot (\delta\tens{\phi}^\mrm{\alpha,P}-\delta\tens{\phi}^\mrm{\alpha,Q} )+ \lb \tens{l}^\alpha \times \tens{F}^\alpha \rb\cdot \delta\tens{\phi}^\mrm{\alpha,Q} = \int_{V_{\mrm c}} \lb \tens{\mu}: \nabla  \delta \tens{\omega} - \tens{S}: \lb\tens{\epsilon}\cdot\delta \tens{\omega} \rb \rb \dif V.
\end{align}
With \eqref{eq:beamvec}, \eqref{eq:rotation_ansatz},  \eqref{eq:P_rot} this can be written as 
\begin{align}
\sum_{\alpha=1}^{N_\mathrm{b}}   \tens{M}^\alpha \cdot \lb \tens{\Phi}\cdot\tens{l}^\alpha \rb
+ \lb \tens{l}^\alpha \times \tens{F}^\alpha \rb\cdot \lb \tens{\Phi}\cdot\tens{x}^\mrm{\alpha,Q}  + \tens{r} \rb = \int_{V_{\mrm c}} \lb \tens{\mu}: \tens{\Phi} - \tens{S}: \tens{\epsilon}\cdot\lb \tens{\Phi}\cdot\tens{x}  + \tens{r} \rb  \rb  \dif V.
\label{eq:virt_rot2}
\end{align}
By using Eq. \eqref{eq:centroid_pos}, the stress-force relation \eqref{eq:stress_force_relation} and the fact that the stresses and $\tens{\Phi}$ are constant in the control volume, the second term of the right hand side of \eqref{eq:virt_rot2}  can be approximated by
\begin{align}
\begin{split}
-\int_{V_{\mrm c}} \tens{S}: \tens{\epsilon}\cdot\lb \tens{\Phi}\cdot\tens{x}  + \tens{r} \rb   \dif V &= -\lb \langle \tens{S} \rangle_\mrm{c} : \tens{\epsilon} \rb \cdot\lb \tens{\Phi} \cdot \tens{x}^\mrm{P}  + \tens{r} \rb \\
&= \sum_{\alpha=1}^{N_\mathrm{b}}  \lb \tens{l}^\alpha \times  \tens{F}^\mrm{\alpha} \rb \cdot \lb \tens{\Phi} \cdot \tens{x}^\mrm{P}  + \tens{r} \rb.
\end{split}
\label{eq:rhs_virt_rot2}
\end{align}
With that \eqref{eq:virt_rot2} can be simplified to
\begin{align}
\sum_{\alpha=1}^{N_\mathrm{b}}   \tens{M}^\alpha \cdot \lb \tens{\Phi}\cdot\tens{l}^\alpha \rb  
+ \lb \tens{l}^\alpha \times \tens{F}^\alpha \rb\cdot \lb \tens{\Phi}\cdot\tens{x}^\mrm{\alpha,Q}  + \tens{r} \rb &= \int_{V_{\mrm c}}\tens{\mu}: \tens{\Phi}  \dif V+\sum_{\alpha=1}^{N_\mathrm{b}}  \lb \tens{l}^\alpha \times  \tens{F}^\mrm{\alpha} \rb \cdot \lb \tens{\Phi} \cdot \tens{x}^\mrm{P}  + \tens{r} \rb,\\\sum_{\alpha=1}^{N_\mathrm{b}}   \tens{M}^\alpha \cdot \lb \tens{\Phi}\cdot\tens{l}^\alpha \rb  
- \lb \tens{l}^\alpha \times \tens{F}^\alpha \rb\cdot \lb \tens{\Phi}\cdot\tens{l}^\mrm{\alpha} \rb &= \int_{V_{\mrm c}}\tens{\mu}: \tens{\Phi}  \dif V .
\end{align}
Again the linear mapping $\tens{\Phi}$ is an arbitrary constant tensor so that we obtain for the averaged couple stress
\begin{align}
\langle \tens{\mu} \rangle_\mrm{c} = \frac{1}{V_{\mrm c}} \int_{V_{\mrm c}}  \tens{\mu} \dif V
&=\frac{1}{V_{\mrm c}} \sum_{\alpha=1}^{N_\mathrm{b}}  \tens{M}^\alpha \otimes \tens{l}^\alpha - \lb \tens{l}^\alpha \times  \tens{F}^\mrm{\alpha} \rb \otimes  \tens{l}^\alpha \\
 &= -\frac{1}{V_{\mrm c}} \sum_{\alpha=1}^{N_\mathrm{b}}  \tens{M}^\mathrm{\alpha,Q} \otimes \tens{l}^\alpha.
\end{align}
The system average is computed in the same way as for the stresses
\begin{align}
\langle \tens{\mu} \rangle_\mrm{S} =  \frac{1}{V_{\mrm S}} \sum_{k=1}^{N_{\mrm c}} V_{\mrm c}^k \langle \tens{\mu} \rangle_\mrm{c}^k.
\end{align}

\subsection{Derivation of equivalent strain fields}
In our derivation we impose the following requirements on the continuous strain and the associated displacement fields:
\begin{itemize}
\item discrete and continuous displacements at the positions of the beam nodes are the same
\item strain fields are constant over control volumes and piecewise constant over the sample domain
\item stress and strain computation is done within identical control volumes 
\end{itemize}

Linear approximations for the spatial dependency of the displacement fields are a natural choice, since they lead to a piecewise constant strain field. For a control volume of arbitrary shape, the averaged strain tensor can be evaluated as
\begin{align}
\langle \tens{e} \rangle_\mrm{c} = \langle \nabla \tens{u} \rangle_{\mrm{c}} = \frac{1}{V_\mrm{c}} \int_{V_\mrm{c}} \nabla \tens{u} \dif V = \frac{1}{V_\mrm{c}} \int_{S_\mrm{c}} \tens{u} \otimes \tens{n} \dif A 
\end{align}
where $\tens{n}$ is the outward pointing normal to the control volume surface $S_{\mrm{c}}$. If the control volume is a $N_\mrm{c}$-sided polygon and the dependency of $\tens{u}$ on the spatial coordinates is linear, we find
\begin{align}
\langle \tens{e} \rangle_{\mrm{c}} = \frac{1}{V_\mrm{c}} \sum_{k=1}^{N_\mrm{c}} \left(\frac{\tens{u}^{k} + \tens{u}^{k+1}}{2} \otimes \tens{n}^{(k,k+1)}\right) \tens{b}^{(k,k+1)}.
\label{eq:strainavs}
\end{align}
Here $\tens{u}_k$ is the displacement of corner node number $k$ in a clockwise enumeration, $\tens{b}^{(k,k+1)}$ is the length of the polygon side connecting nodes $k$ and $k+1$, and $\tens{n}^{(k,k+1)}$ the outward pointing normal vector to this side. Note, that similar results have been obtained by  \cite{bagi1996} for an assembly of granules.
Applying Eq.~\eqref{eq:strainavs} to the control volumes, the displacements of the corners of the control volume are only partially known: While the displacements of the corners that are located at the midpoint of a beam are explicitly known, they are unknown for the corners located at the cell midpoints. 
Owing to the assumption of constant strain/stress any sub-volume of the control volume must have the same (constant) strain field as the whole control volume. Therefore we approximate the control volume strain by computing the strain from the beam midpoints for which the displacements are known.
For the calculation of the continuum rotation gradient $\langle \nabla \tens{\omega} \rangle_\mrm{c}$ we follow the same argument for beam rotations (instead of beam displacements). To consistently compute the averaged rotation $\langle \tens{\omega} \rangle_\mrm{c}$ a linear interpolation for the polygon centroid from the same known beam midpoints is used. With the computed averaged quantities the averaged strain measures can be expressed as
\begin{align}
\langle \tens{\varepsilon} \rangle_\mrm{c} &= \frac{1}{2} \lb \langle \tens{e}  \rangle_{\mrm{c}} + \langle \tens{e}  \rangle_{\mrm{c}}\T \rb, \\
\langle \tens{\varepsilon}^\mrm{R} \rangle_\mrm{c} &= \frac{1}{2} \lb \langle \tens{e} \rangle_{\mrm{c}} - \langle \tens{e}  \rangle_{\mrm{c}}\T \rb + \tens{\epsilon}\cdot \langle \tens{\omega} \rangle_\mrm{c}.
\end{align} 

\subsection{Verification of the continuization method}
\begin{figure}
\centering
\begin{minipage}[c]{0.85\textwidth}
\subcaptionbox{$\beta= 0.0$}
{\includegraphics[width=0.325\textwidth]{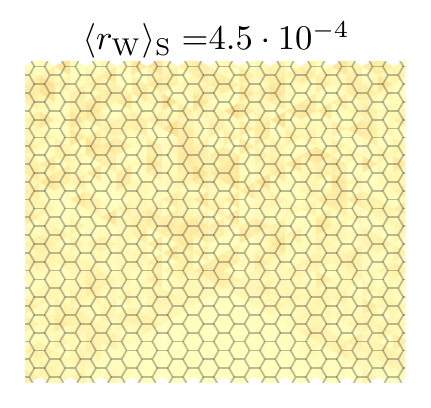}}
\subcaptionbox{$\beta= 0.1$}
{\includegraphics[width=0.325\textwidth]{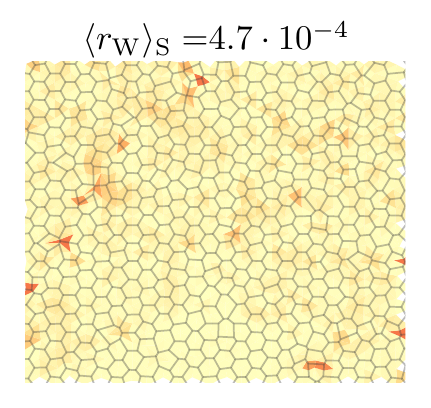}}
\subcaptionbox{$\beta= 5.0$ \label{fig:energies_c}} 
{\includegraphics[width=0.325\textwidth]{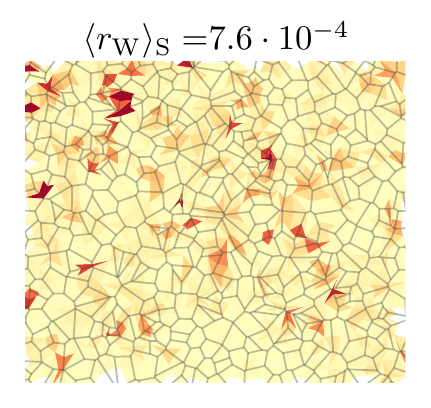}}
\end{minipage}
\begin{minipage}[c]{0.115\textwidth}
\vspace{-0.8em}
{\includegraphics[width=1.0\textwidth]{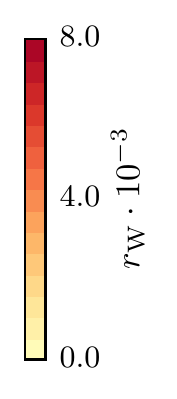}}
\end{minipage}
\caption{Energy residuals for different microstructures with fixed system size $H=19 \Delta p$ under uniaxial compressive loading.}\label{fig:energies}
\end{figure}

As a first verification of the continuization procedure we test for the energy consistency 
by comparing for each control volume the continuum strain energy $W_\mrm{c}$ with the energy stored in the beams $W_\mrm{b,c}$. In analogy  to \Secref{sec:beamenergy} the elastic energy of a a discrete beam is given by
\begin{align}
W_\mrm{b} = 
\frac{1}{2} \left(\tens{F}^\mrm{P} \cdot\tens{w}^\mrm{P} + \tens{F}^\mrm{Q}  \cdot\tens{w}^\mrm{Q} 
+\tens{M}^\mrm{P}\cdot\tens{\phi}^\mrm{P}+\tens{M}^\mrm{Q}\cdot \tens{\phi}^\mrm{Q} \right) 
\end{align}
with forces $ \tens{F}^\mrm{P}, \tens{F}^\mrm{Q}$, moments $\tens{M}^\mrm{P}, \tens{M}^\mrm{Q} $, displacements $\tens{w}^\mrm{P}, \tens{w}^\mrm{Q}$ and rotations $\tens{\phi}^\mrm{P}, \tens{\phi}^\mrm{Q}$ given at the beam end points P and Q. Note, that P corresponds to the junction point and Q to the midpoint between two junctions. The total beam energy is then given by the sum of energies of all beams in the control volume $V_\mrm{c}$:
\begin{align}
W_\mrm{b,c} = \sum_{\alpha=1}^{N_\mrm{b}} W_\mrm{b}^\alpha.
\end{align}
Because stresses and strains are by construction constant in the control volumes, the continuum strain energy can be computed as
\begin{align}
\label{eq:strain_energy}
W_\mrm{c} = 
\frac{1}{2} \lb \langle\tens{\sigma}\rangle_\mrm{c} : \langle \tens{\varepsilon} \rangle_\mrm{c}  + \langle \tens{S}\rangle_\mrm{c}  :\langle \tens{\varepsilon}^\mrm{R} \rangle_\mrm{c}
+ \langle \tens{\mu} \rangle_\mrm{c} : \langle \nabla \tens{\omega} \rangle_\mrm{c}\rb.
\end{align}
As a measure for the error the relative energy residual $r_\mrm{W} = {| W_\mrm{c}-W_\mrm{b,c} | }/{W_\mrm{b,c}}$
is used, which is shown in \figref{fig:energies} for three structures of increasing randomness. For the regular structure $r_\mrm{W}$ is almost equally distributed, with the trend of smaller errors towards less displaced beams. Increasing randomness leads to an increase in local errors as the volumes become more distorted and the centroid approximation, \eqref{eq:centroid_pos}, is violated. Distortion of the control volumes alone, however, does not necessarily give rise to larger errors, because the violation of the centroid approximation becomes relevant only for rotation dominated energies. Nonetheless, even for the highly irregular system the maximum error is smaller than $1\%$ and the weighted average of the global error is for all systems less than $0.1\% $ (compare the highly distorted control volumes in \figref{fig:energies_c}, of which only some show a large error). 
We conclude that the chosen approximations are sufficiently accurate and able to represent rotational degrees of freedoms as opposed to a classical Cauchy continuum (see~\ref{appendix:cauchy formulation}).

\begin{figure}
\centering {Stresses}
\begin{minipage}[c]{1.\textwidth}
{\includegraphics[width=.235\textwidth]{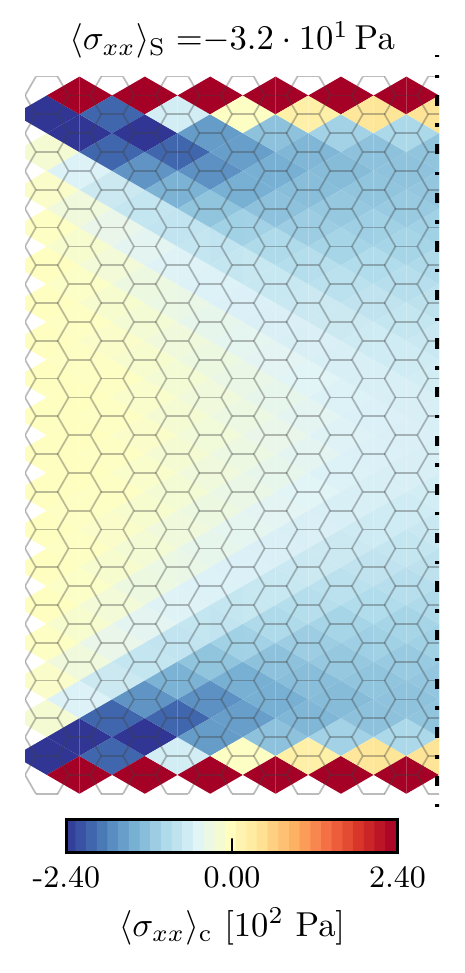}}
{\includegraphics[width=.235\textwidth]{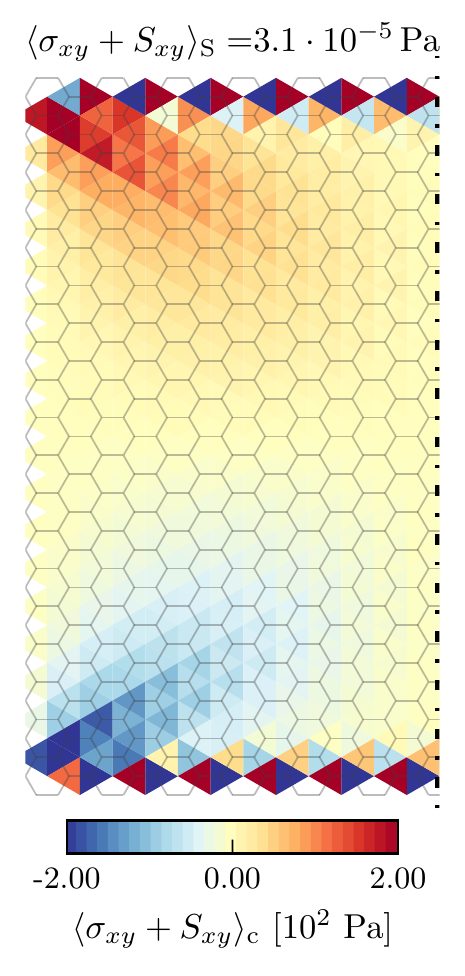}}
{\includegraphics[width=.235\textwidth]{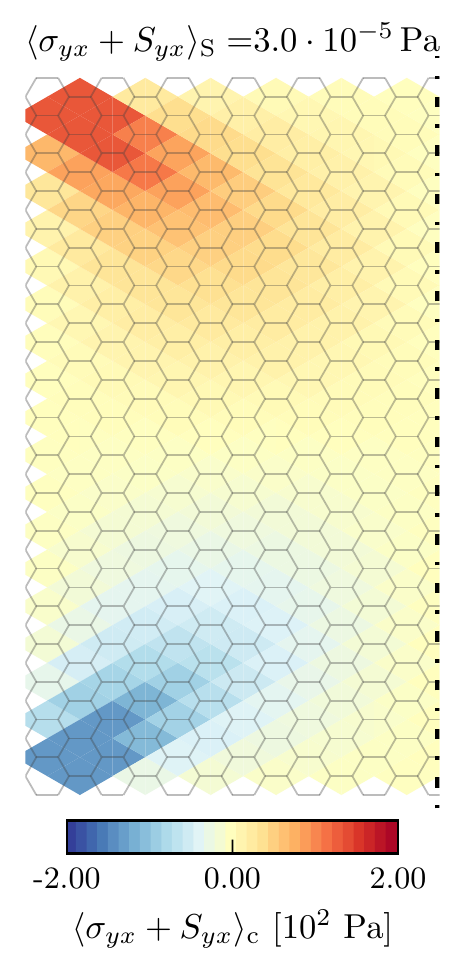}}
{\includegraphics[width=.235\textwidth]{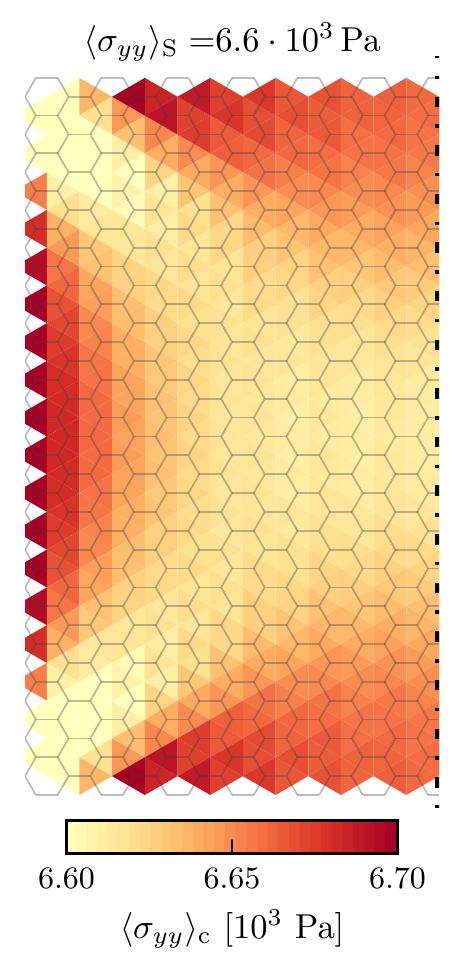} }
\end{minipage}
\centering {Strains}
\begin{minipage}[c]{1.\textwidth}
{\includegraphics[width=.235\textwidth]{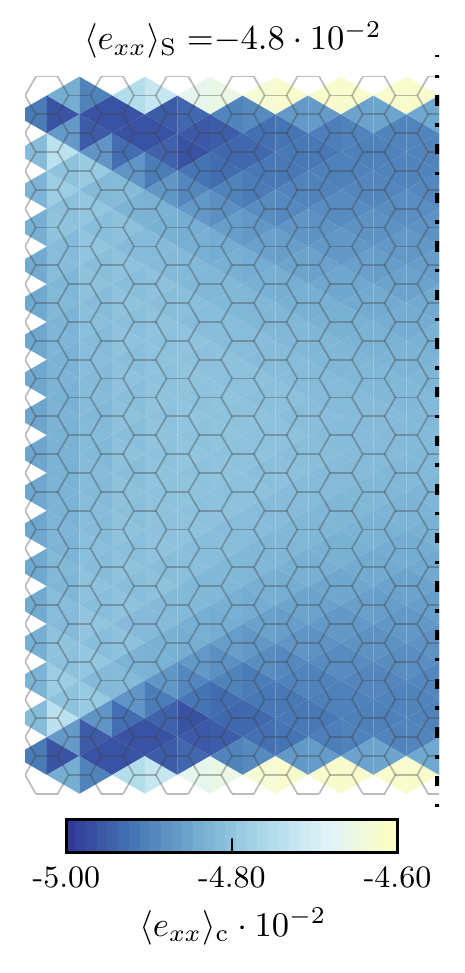}}
{\includegraphics[width=.235\textwidth]{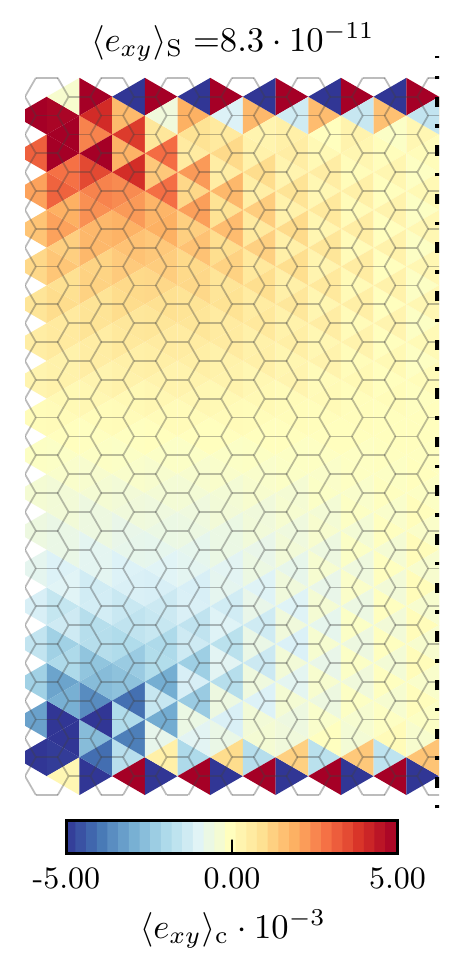}}
{\includegraphics[width=.235\textwidth]{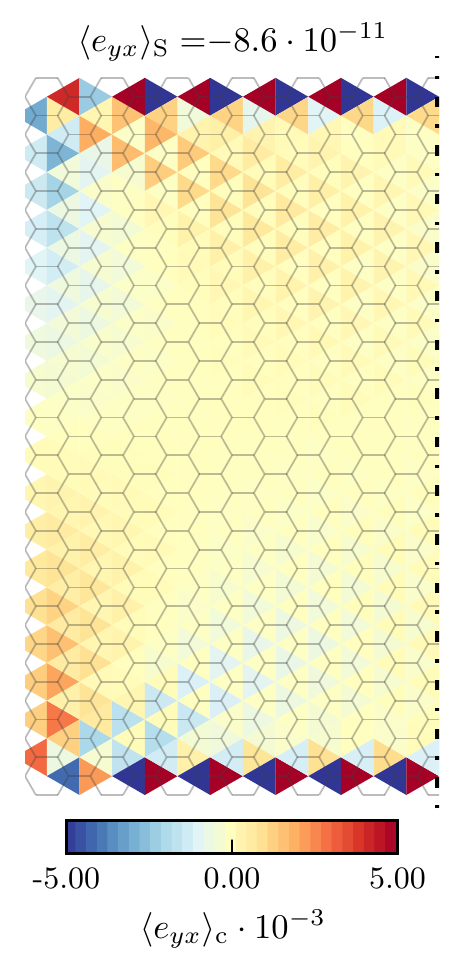}}
{\includegraphics[width=.235\textwidth]{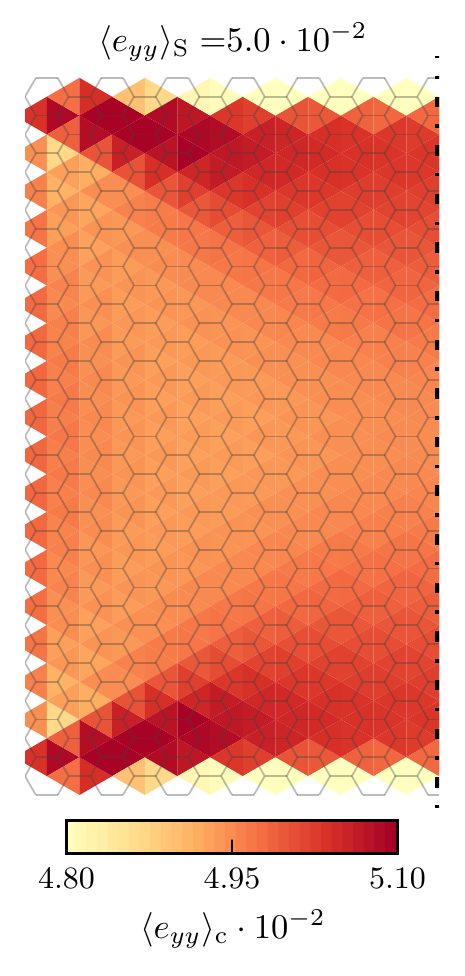} }
\end{minipage}
\caption{Stress and strain tensor components for a regular honeycomb structure ($\beta=0,\, H=19 \Delta p$) with closed boundary cells under uniaxial compression. Note that only the left half of the system is shown as it is symmetric with respect to its central axis. \label{fig:reg_stress_strain}}
\end{figure}

As an other verification of the continuization method we analyze the local stresses and strains. As reference system we chose a regular system under uniaxial compression (cf. \figref{fig:reg_stress_strain}). We observe that both stresses and strains show a similar inhomogeneous distribution. The normal stresses $\langle \sigma_{yy}\rangle_\mrm{c}$ are dominating, with a weighted average of $ \langle \sigma_{yy}\rangle_\mrm{S} = 6.6\cdot 10^3$ Pa which is close to the value of $\sigma_{yy}^* = 6.86\cdot 10^3$ Pa obtained by analysis of the reaction forces. Furthermore one can see that boundary effects are present near those boundaries where the system is constrained. Near these boundaries we observe checker-board patterns of shear stress and shear strain which are artifacts of the continuization method. The averaged strain $\langle e_{yy}\rangle_\mrm{S}$ is in good agreement with the imposed engineering strain $e^*=0.05$. The lateral compression is slightly smaller than the axial dilatation, which results in an averaged 2D Poisson ratio of
\begin{align}
\left \langle \nu \right \rangle_\mrm{S} = - \left \langle \frac{e_{yy}}{e_{xx}} \right \rangle_\mrm{S} \approx 0.97
 \end{align}
which is close to the theoretical value for regular honeycombs, cf. Eq. \eqref{eq:theo_vals} in \Secref{sec:glob_results}. 

To conclude, our continuization method is not only energy consistent with the discrete system, but also gives comparable results on the macroscopic scale for stresses and strains. Discretization/continuization artifacts such as artificial stress or strain oscillations are largely absent in the bulk of the sample but may be present in the immediate vicinity of constrained boundaries.

\section{Results: Size and disorder-dependent elasticity of foam microstructures}
\label{sec:results}
\begin{figure}[htp]
\centering
\begin{subfigure}[b]{0.245\textwidth}
{\includegraphics[width=1\textwidth]{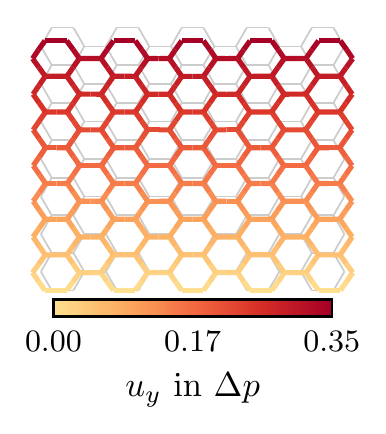}}
\caption{Compression, $\beta=0$}
\end{subfigure}
\begin{subfigure}[b]{0.245\textwidth}
{\includegraphics[width=1\textwidth]{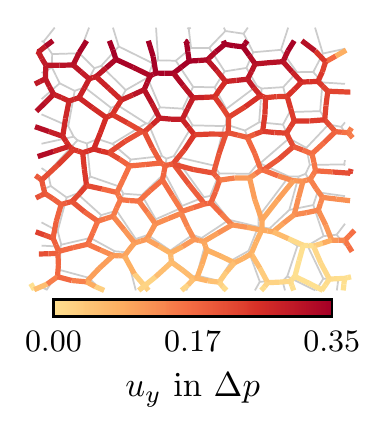}}
\caption{Compression, $\beta=0.3$}
\end{subfigure}
\begin{subfigure}[b]{0.245\textwidth}
{\includegraphics[width=1\textwidth]{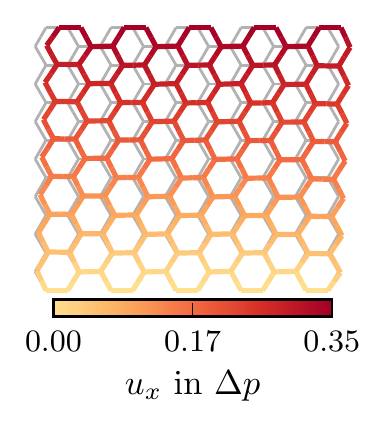}}
\caption{Simple shear, {$\beta=0$}}
\end{subfigure}
\begin{subfigure}[b]{0.245\textwidth}
{\includegraphics[width=1\textwidth]{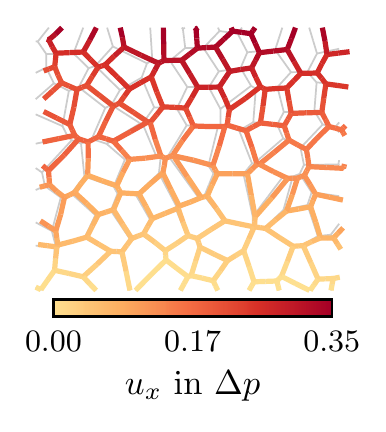}}
\caption{Simple shear, $\beta=0.3$}
\end{subfigure}
\caption{Undeformed (in grey) and deformed structures for uniaxial compression along the $y$ direction and for simple shear loading; disorder parameters $\beta=0$ and $\beta=0.3$; the system height is for all systems $H=7 \Delta p$. The colorscale represents the local displacement in $y$ direction for compression and in $x$ direction for shear loading.}\label{fig:disps}
\end{figure}

In the following we study the behavior of regular and random microstructures constructed as discussed in \Secref{sec:model} under two loading conditions, simple shear and uniaxial compression, described in \Secref{sec:BC} and Table~\ref{tb:Dirichlet BC}. As mentioned in \Secref{sec:BC} the simulations are performed with the aspect ratio $W/H=R_0=2/\sqrt{3}$ if not stated otherwise. The deformation of the systems is shown for a regular ($\beta=0$) and one realization of a random $(\beta=0.3)$ system in  \figref{fig:disps}. 
It can be seen that in the regular system the beam and junction displacements show an almost continuous, linear distribution. The irregular system on the other hand exhibits a more inhomogeneous distribution of the displacements, corresponding to inhomogeneities in the local strain fields.

\subsection{Macroscopic system response}
\label{sec:glob_results}
The macroscopic system response (cf. \Secref{sec:sys_response}) is influenced by aspect ratio, system size,  irregularity of the microstructure, and configuration of the microstructure at the boundaries. In what follows we compare the relevance of these parameters for systems of different size. Regular honeycomb structures are chosen as reference systems since their deformation properties can be considered isotropic on the macroscopic scale (\cite{silva1995}, \cite{gibson1999}). Thus, the stiffnesses \eqref{eq:macro_stiffness} can be interpreted as a geometry dependent Young's Modulus $E^*$ for uniaxial compression and shear modulus $G^*$  for pure shear. As an analytical benchmark the solutions of \cite{gibson1999} are chosen. These solutions take beam bending, axial and shear deformation into account and give for an 
infinite system and the material properties $\rho_{\mrm{rel}}=0.1, E_{\mrm{B}}=1\cdot 10^8\mrm  \, \mrm{Pa}, $ and $\nu_{\mrm{B}}=0.3$ 
as used in our simulations the theoretical system parameters
\begin{align}
\label{eq:theo_vals}
 E^*_{\mrm T} &\approx 1.44 \cdot 10^5 \, \mrm{Pa}, & \nu^*_{\mrm T} &\approx 0.971,& G^*_{\mrm T} &\approx 3.65 \cdot 10^4  \, \mrm{Pa}. 
\end{align}
In the following, the values of the macroscopic response parameters $C^*_{\mrm c,s}$ are always normalized by the theoretical Young's modulus $ E^*_{\mrm T} $ for compression and the theoretical shear modulus $G^*_{\mrm T} $ for shear.

\begin{figure}[htb]
\centering
\hspace*{\fill}%
{\includegraphics{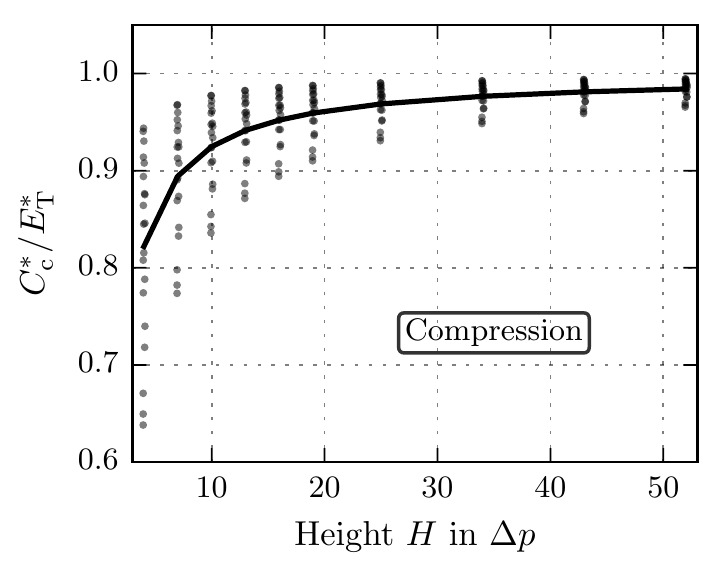}}
\hfill
{\includegraphics{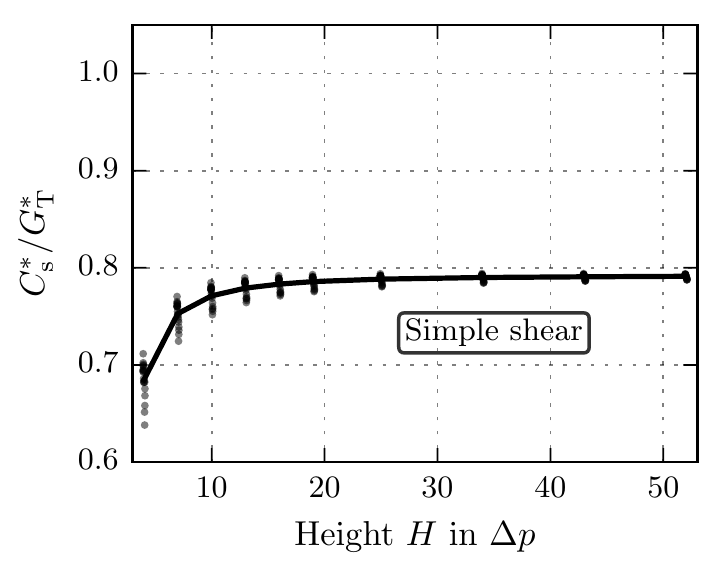}}
\hspace*{\fill}%
\caption{Normalized system responses evaluated for multiple cut-outs from a regular honeycomb structure ($\beta=0$). Each dot represents one simulation, and the average of all cut-outs is represented as the black line.}\label{fig:subrealization}
\end{figure}

\paragraph{Influence of microstructure configuration at the boundary $(\beta=0)$} 
As described in \Secref{sec:BC} varying random cut-outs have a large influence on the size dependency of the mechanical response. This is primarily important for regular or slightly perturbed systems whereas this effect cancels out for systems with a large degree of randomness since the boundaries average over many different cell configurations. 
 \figref{fig:subrealization} reveals that the stiffness values of different cut-outs from a regular honeycomb structure exhibit significant scatter, which decreases from about 30 \% for the smallest to 3 \% for the largest investigated systems, and which is expected to reach zero in the infinite-system limit. Note, that for each investigated system size the realizations slightly vary in size as a result of the scaling procedure (see  \Secref{sec:microproperties}).
Thus, the average stiffness is obtained by first sorting all cut-outs into bins w.r.t. their original, unscaled size. Then all values corresponding to each size/bin are averaged.
Smaller systems are on average weaker than larger ones for both loading conditions. With increasing system size the stiffness approaches asymptotically the bulk value. For uniaxial compression this limit is close to the theoretical Young's modulus $E^*_{\mrm T}$, whereas for simple shear loading the asymptotic parameter $C^*_{\mrm s}$ remains significantly below the theoretical shear modulus $G^*_{\mrm T}$. This is expected since simple shear loading does not lead to a uniaxial macroscopic stress and strain state. In fact, the asymptotic value of the scaled parameter $C^*_{\mrm s}$ approached in our simulation matches well the result obtained for an isotropic material with the same aspect ratio, calculated with the theoretical bulk values of the elastic moduli. For compressive loading one can see that the size effects are much smaller for the stiffest structures compared to the weaker ones. However, this effect is less pronounced for shear loading. 

\begin{figure}
\centering
\hfill
{\includegraphics{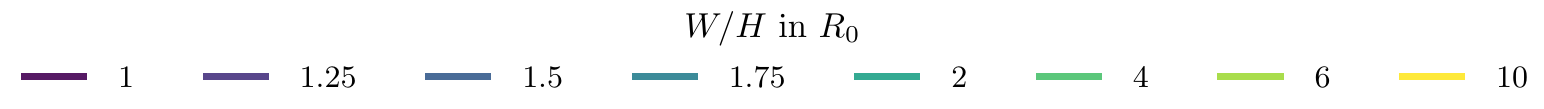}} 
\hfill\vspace{-0.5em}
\hspace*{\fill}
{\includegraphics{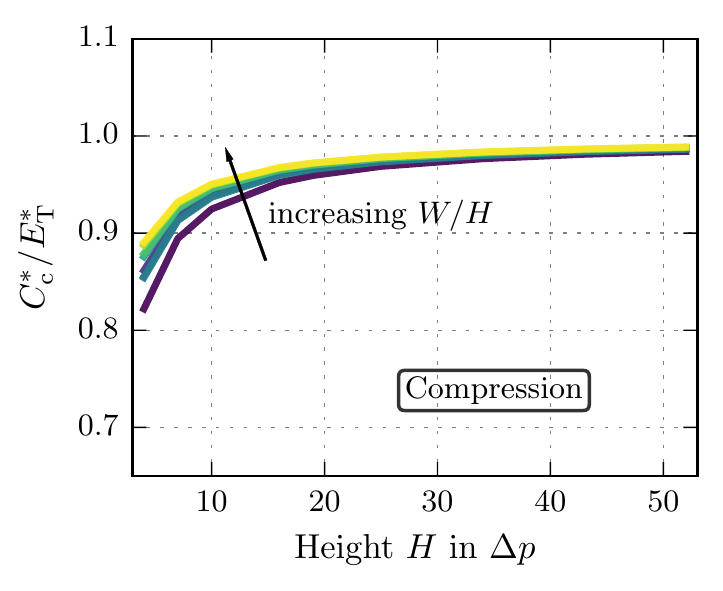}}
\hfill
{\includegraphics{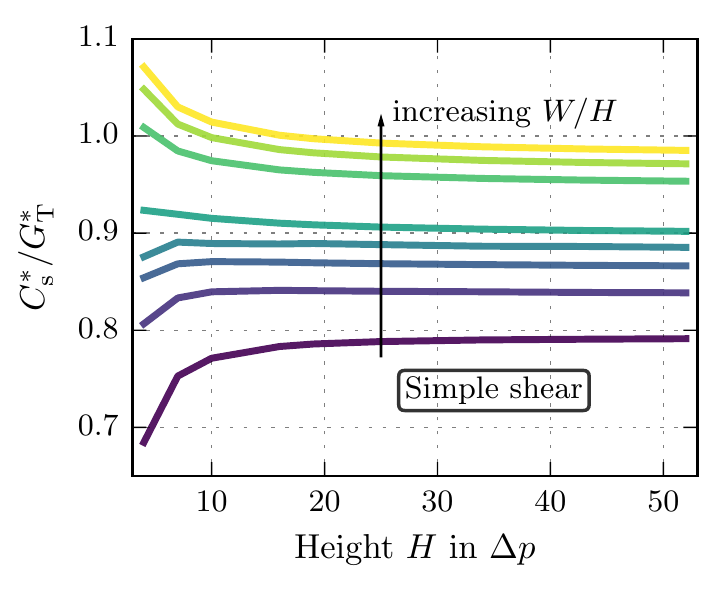}}
\hspace*{\fill}%
\caption{Normalized system responses for regular systems ($\beta=0$) of different aspect ratios.}\label{fig:aspect_ratio}
\end{figure}

\begin{figure}
\centering
\hfill
{\includegraphics{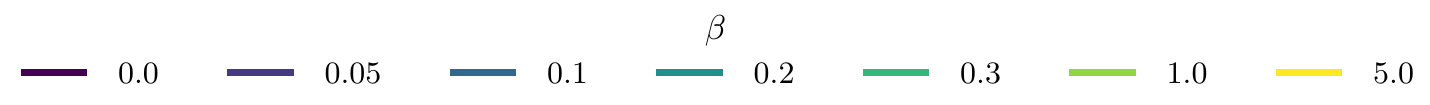}} 
\hfill\vspace{-0.5em}
\hspace*{\fill}
{\includegraphics{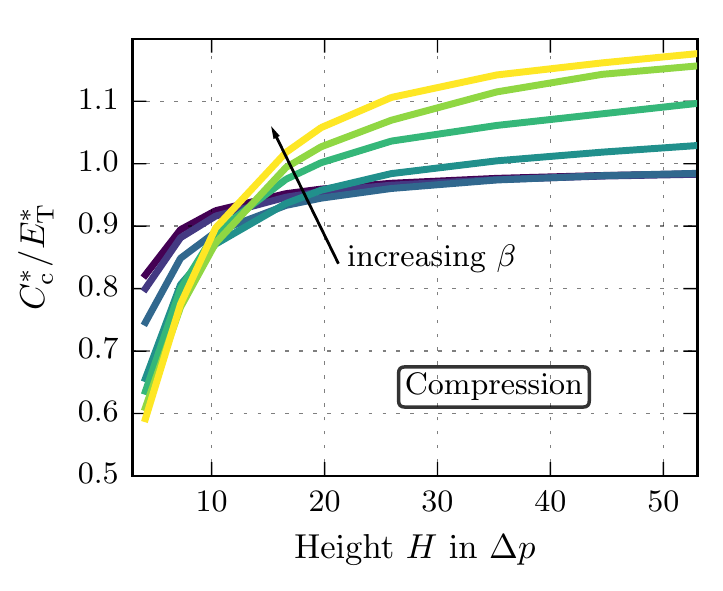}}
\hfill
{\includegraphics{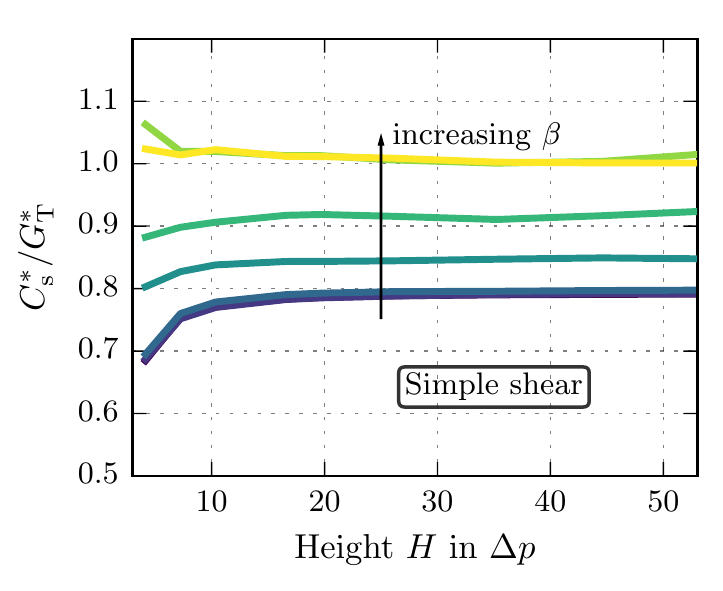}}
\hspace*{\fill}
\caption{Normalized global system response for uniaxial compression and simple shear loading for different degree of randomness.} 
\label{fig:size_effects}
\end{figure}

\paragraph{Influence of system aspect ratio on elastic response $(\beta=0)$} Our second investigation concerns the effect of aspect ratio $R=W/H$ on the mechanical response. For this analysis, we vary for each system height the system width in such a manner that the resulting aspect ratios are multiples of the aspect ratio of a single regular honeycomb $R_0=2/\sqrt{3} \approx 1.155$. Results are shown in \figref{fig:aspect_ratio}.
Again, one observes a clear "smaller is weaker" trend. In compression, the stiffness for each height increases with increasing $W$ but this increase soon saturates and even for very wide systems a strong dependency of stiffness on system height remains. In simple shear, by contrast, the size effect becomes smaller for wider systems and eventually inverts to a "smaller is stiffer" behavior. Moreover, for very wide and high systems the shear stiffness approaches the shear modulus of a bulk system, $C^*_{\mrm s} \to G^*_{\mrm T}$, as stress multi-axialities at the specimen corners and stress-free regions near the specimen side boundaries become asymptotically irrelevant. Again the behavior of large systems is consistent with that of an isotropic reference material without microstructure. 
\paragraph{Influence of microstructure randomness on system response $(\beta\neq 0)$}
To determine statistically representative stiffness parameters for microstructurally disordered systems, averages over different statistical realizations of any given microstructure must be considered. In general, bigger system sizes reduce the influence of specific microstructural features as they contain a wider spectrum of local configurations, leading to reduced fluctuations of the system-scale stiffness.  Accordingly, we average over 200 different realizations for systems with $H \leq 16 \Delta p$ and 100 different realizations for $H > 16 \Delta p $. For each realization of a disordered microstructure again 20 cut-outs are taken. The resulting average response coefficients  are shown for different values of $\beta$ in  \figref{fig:size_effects} as functions of system size. 

The overall picture is complex, and for interpreting the observations it is important to note that the elastic moduli of bulk systems actually {\em increase} with increasing disorder as pointed out by \cite{zhu2001}.  Accordingly, for large systems the macroscopic response coefficients in our simulations increase with increasing $\beta$. At the same time, also irregular systems exhibit size effects though this size-dependent behavior depends on the loading mode. It is a pronounced "smaller is weaker" behavior in compression where disorder exacerbates the weakening effect. As a consequence, in compression for the smallest systems the dependency of stiffness on disorder is reversed (the systems with the largest $\beta$ are weakest). In simple shear, on the other hand, the size dependency of stiffness is much weaker.
 Here, a complex picture emerges where the size effects invert for increasing degree of disorder. Again this needs to be interpreted in view of the corresponding bulk behaviour: Under shear loading, the individual beams of a honeycomb structure deform primarily by bending. As investigated by \cite{hyun2002}, microstructures which allow for more stretch/compression dominated beam deformations, such as triangular or so-called Kagome structures, are significantly stiffer than regular honeycombs. Irregularities affect the morphology of the local networks by introducing polygons with reduced number of edges, or by shortening some edges to an extent that the cells resemble such polygons. 

\begin{figure}[htb]
\centering
\hspace*{\fill}
\subcaptionbox{Perturbed seeds in the center.}
{\includegraphics{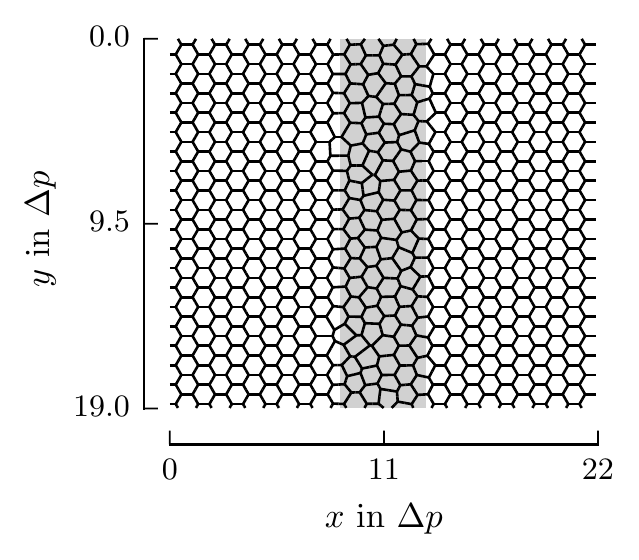}}
\hfill
\subcaptionbox{Perturbed seeds at the boundaries.}
{\includegraphics{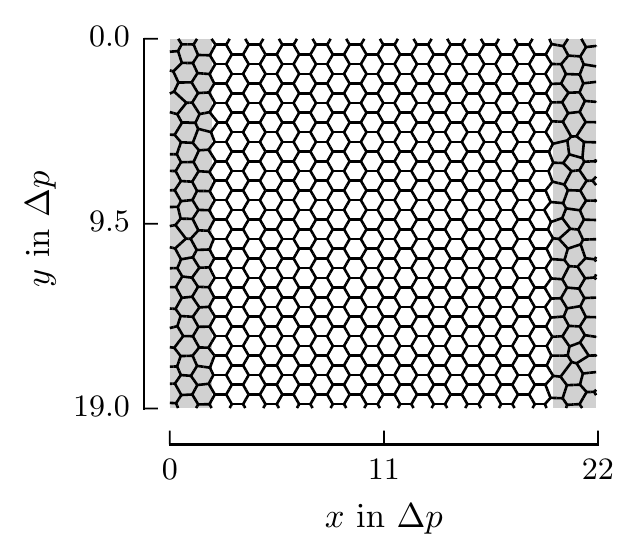}}
\hspace*{\fill}
\caption{Examples of locally perturbed microstructures with Voronoi seeds for $H=19 \Delta$. $20 \%$ of the seeds are perturbed either (a) within a central vertical band or (b) near the left and right boundaries. The shaded areas indicate the zones in which the seeds are perturbed.}  \label{fig:perturbated_microstructures}
\end{figure}

\begin{figure}
\centering
\hfill{\includegraphics{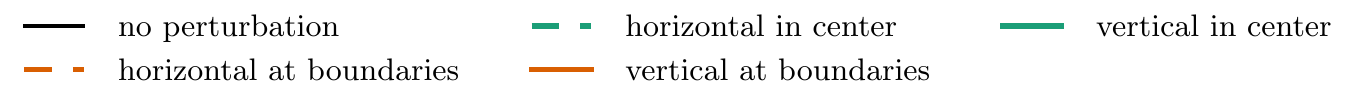}}\hfill\vspace{-0.5em}
\hspace*{\fill}
{\includegraphics{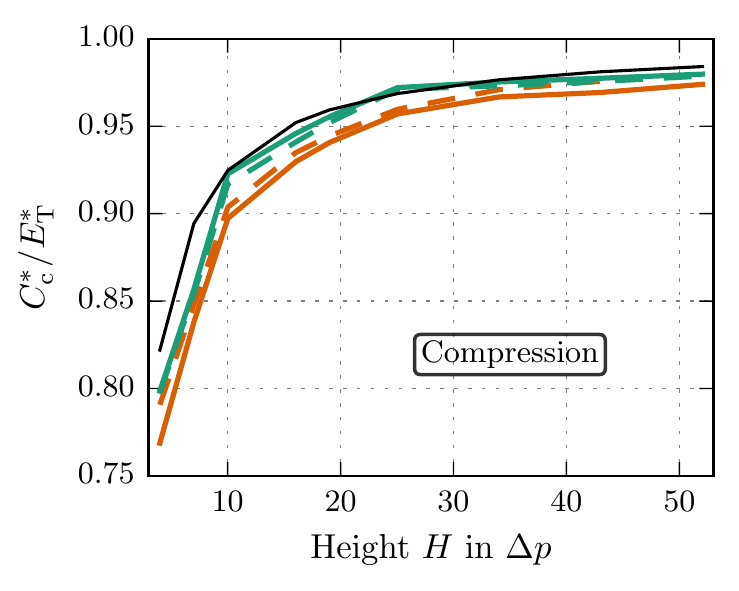}}
\hfill
{\includegraphics{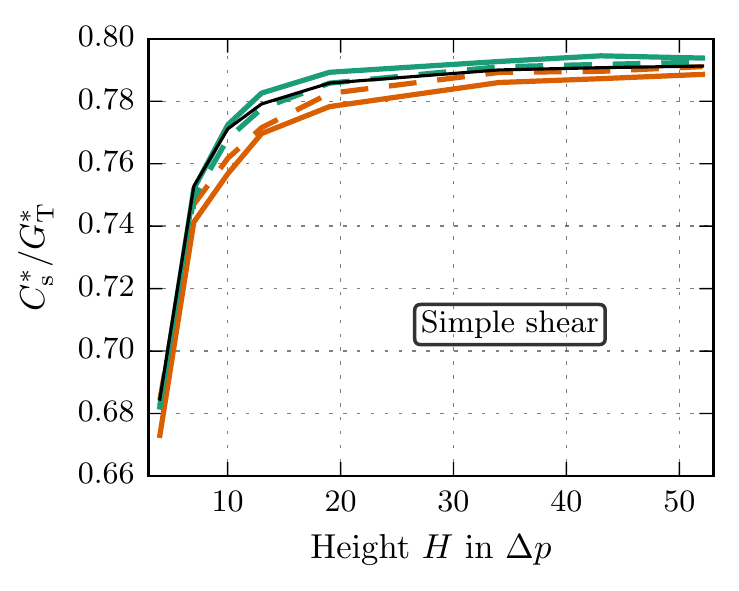}}
\hspace*{\fill}
\caption{Comparison of system responses in uniaxial compression and simple shear for the four types of 
locally perturbed microstructures.} 
\label{fig:sys_response_perturbated_microstructures}
\end{figure}

\paragraph{Influence of the spatial distribution of microstructural irregularities} To investigate possible influences of the spatial distribution of disorder, we analyze microstructures where the regular honeycomb pattern has been perturbed by displacement of seeds only in parts of the domain. Specifically, we investigate what happens if perturbations are localized near the sample boundaries or conversely in the center. Starting from a regular seeding, $20 \%$ of the seeds are perturbed with a perturbation factor of $\beta=0.1$. Four different perturbation patterns are considered:
(i) vertical bands at the left and right ($0<x_{\cal S}<0.1 W$ or  $0.9 W < x_{\cal S} < W $),
(ii) a vertical band in the center of the system ($0.4 W < x_{\cal S} < 0.6 W$),
(iii) horizontal bands at the top and bottom ($0<y_{\cal S}<0.1 H$ or  $0.9 H <y_{\cal S}< H $) and
(iv) a horizontal band in the center of the system ($0.4 H <y_{\cal S}< 0.6 H$),
where $x_{\cal S}, y_{\cal S}$ are the coordinates of the seeds ${\cal S}$ of the Voronoi tessellation. For each case 100 different random perturbations with 20 random cut-outs are taken. As an example, \figref{fig:perturbated_microstructures} shows two realizations with vertical perturbation bands for a system of size $H=19 \Delta p$. 
The resulting system responses, together with that of a regular system as reference, are depicted in \figref{fig:sys_response_perturbated_microstructures}. One can see that weakest are microstructures where perturbations are located along the free boundaries of the sample. A vertical perturbation band in the center increases slightly the stiffness in simple shear whereas the perturbations at the boundaries reduce the macroscopic stiffness. This finding indicates that softening associated with microstructure perturbations may be more pronounced near free surfaces.

\subsection{Microstructural analysis}
\begin{figure}
\centering
\begin{minipage}[c]{0.88\textwidth}
\centering
{\rotatebox{90}{\small $\beta=0$}}
\begin{minipage}[c]{0.43\textwidth}
\centering
{\small Compression}
{\includegraphics{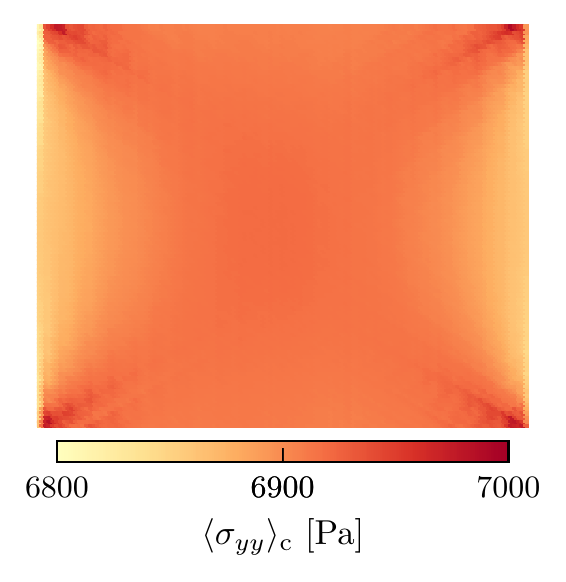}}
\end{minipage}
\begin{minipage}[c]{0.43\textwidth}
\centering
{\small Simple Shear}
{\includegraphics{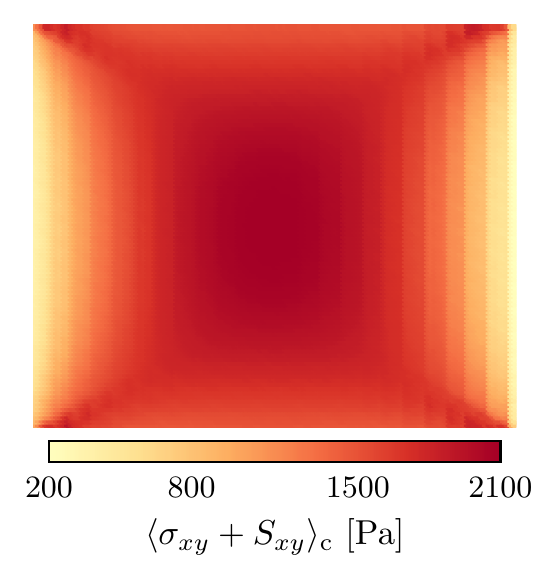}}
\end{minipage}
\end{minipage}
\begin{minipage}[c]{0.88\textwidth}
\centering
{\rotatebox{90}{\small $\beta=0.3$}}
\begin{minipage}[c]{0.43\textwidth}
\centering
{\includegraphics{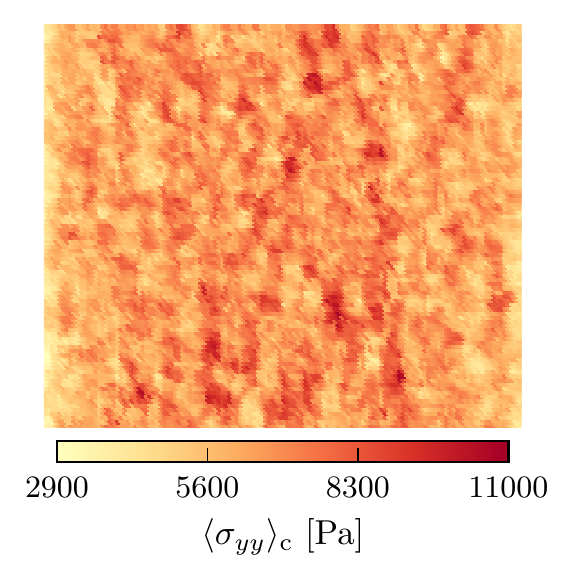}}
\end{minipage}
\begin{minipage}[c]{0.43\textwidth}
\centering
{\includegraphics{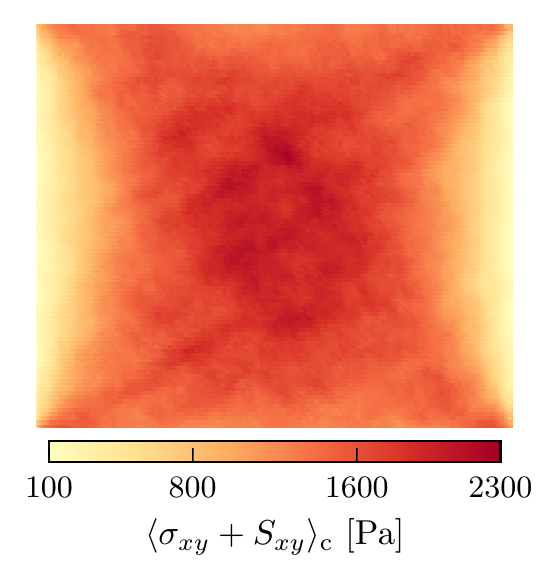}}
\end{minipage}
\end{minipage}
\caption{Ensemble averaged stresses for uniaxial compression (left) and simple shear loading (right) for regular (top) and random microstructures (bottom), system height $H= 19 \Delta p$. }\label{fig:avg_stress}
\end{figure}

\begin{figure}
\centering
\hfill
{\includegraphics{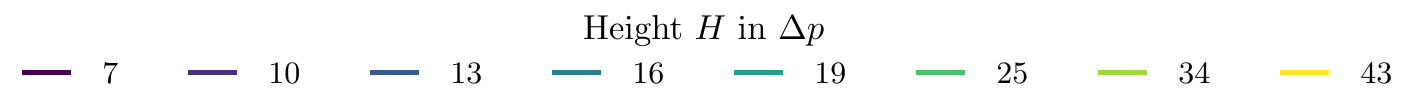}}
\hfill\vspace{-0.5em}
\hspace*{\fill}
\begin{subfigure}[c]{.48\textwidth}
{\includegraphics{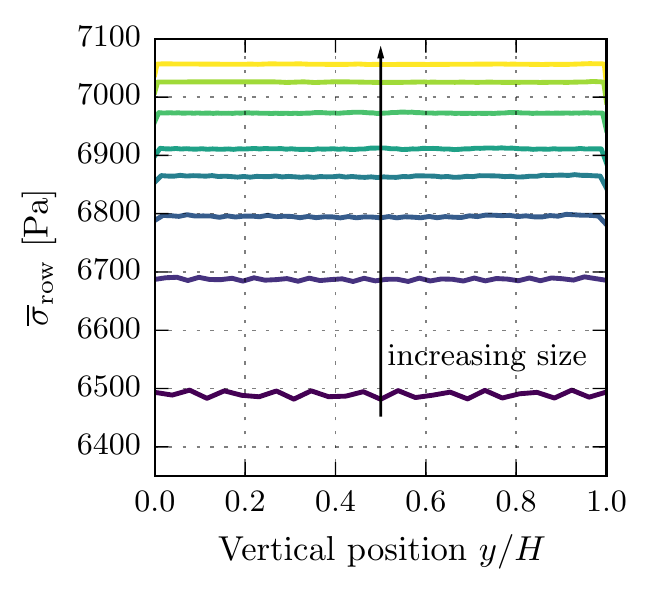}}
\caption{stress profiles parallel to stress axis}\label{fig:stress_profiles_row}
\end{subfigure}
\hfill
\begin{subfigure}[c]{.48\textwidth}
{\includegraphics{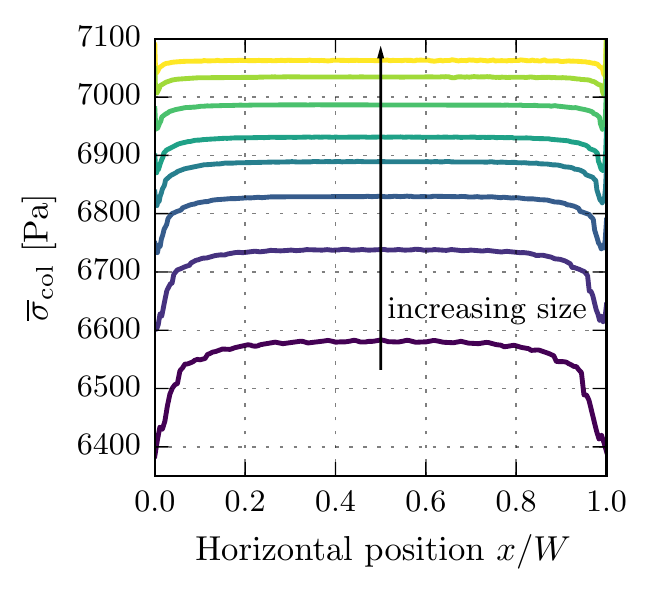}}
\caption{stress profiles perpendicular to stress axis}\label{fig:stress_profiles_col}
\end{subfigure}
\caption{Averaged stress profiles for uniaxial compression of regular honeycomb structures of different sizes for the }\label{fig:stress_profiles}
\end{figure}

We now turn to analyzing random microstructures in terms of stress and strain patterns. To quantify these patterns we use the continuization method presented in \Secref{sec:homogenization}. In order to obtain an averaged stress pattern from different cut-outs and microstructure realizations a common discretization is needed. To that end the stress and strain values are mapped by nearest-neighbor interpolation onto a regular grid which is the same for each system size and regularity factor. By choosing the grid spacing such that there are at least 2--3 grid points in each Voronoi cell, the error introduced by interpolation remains small and local deformation patterns are preserved. In a second step an ensemble average over all $N_{\mrm S}$ microstructure realizations, which themselves are averaged over a set of $N_\mrm{CO}$ random cut-outs, is computed. Because the continuization method is not suitable for the outermost beams and may produce artifacts near boundaries, a layer of width $a=\Delta p/ \sqrt{3}$ at the top and bottom boundaries, and of width $r = \Delta p/2$ at the left and right boundary, is excluded from the analysis. As an example, average stresses for systems of size $H =19 \Delta p$ and disorder parameters $\beta = 0$ and $\beta=0.3$ are shown in \figref{fig:avg_stress}. It can be seen that despite the fact that the investigated systems are not continuous, the average response resembles a continuum stress distribution. For uniaxial compression the regular system  shows an almost homogeneous stress distribution. The outermost layers near the free boundary (left and right edge) experience about 2\% lower than average stresses, whereas in the corners slightly higher stresses are observed. The irregular system shows  higher fluctuations in the local stress response with a slight preference for regions of enhanced or reduced stress to mutually align in direction of the loaded stress axis. The results for simple shear show a characteristic deformation pattern with very low stress values near the free boundary, which is again similar to the findings for a homogeneous continuum. In contrast to the compressive case the fluctuations of the random system are without any preferred directionality.
While the inhomogeneous stress distribution in simple shear is mainly dictated by the boundary conditions, for uniaxial compression we would expect a completely homogeneous distribution of stress in a system without microstructure. Conversely, any stress inhomogeneities observed under compressive loading can be directly related to microstructure effects. We thus study the stress distributions emerging under compression by determining, for systems of different size, stress profiles parallel and perpendicular to the compression axis. To this end we perform row- and column-wise averages of the axial stresses: 
\begin{align}
\overline{\sigma}_\mrm{row} &=  \frac{1}{N_\mrm{S} N_\mrm{CO} } \sum^{N_\mrm{S}} \sum^{N_\mrm{CO}} \frac{1}{W-2 a} \int_a^{W-a} \langle \sigma_{yy} \rangle_\mrm{c}\dif x, \\
\overline{\sigma}_\mrm{col} &= \frac{1}{N_\mrm{S} N_\mrm{CO} }\sum^{N_\mrm{S}} \sum^{N_\mrm{CO}} \frac{1}{H-2 r} \int_r^{H-r} \langle \sigma_{yy} \rangle_\mrm{c} \dif y.
\end{align} 
The resulting horizontal and vertical stress profiles are shown in \figref{fig:stress_profiles}. It can be seen that the averaged stresses for bigger systems are higher and saturate towards a limit value, which is in line with the results of the global stiffness response. Profiles along the stress axis show an approximately constant stress level, whereas profiles perpendicular to the stress axis show a reduced stress level near the free surfaces and a stress plateau in the central region of the specimen. For smaller sizes this plateau occupies a smaller fraction of the specimen cross-section and the difference in stress level between the boundary and the central region is more pronounced.

\section{Discussion}
\label{sec:discussion}
The elastic behavior of small samples of open-cellular solids (i.e. $H<100 \Delta p$) is, in addition to a density dependency, strongly affected by size effects and microstructure irregularities. The size-dependent behavior arises from a complex interplay of bulk and surface effects. Since the surface intersects some of the load-carrying beams, it is natural to assume that a near-surface region with a width of about one cell is less stiff than the bulk, and that this might lead to reduced stiffness of smaller samples where surface effects are more relevant. This is consistent with the observation in \figref{fig:stress_profiles} which suggests that, even in compression, stresses are reduced in a near-surface region which in smaller samples occupies a larger volume fraction. However, the same figure also demonstrates that the main reason for the stiffening that occurs with increasing system size is not a reduced importance of the softer boundary layer but rather an increased average stiffness in the {\em sample interior}. 

In our investigation the observed "smaller is weaker" behaviour was studied for systems with non-periodic boundary conditions only. The results are comparable to those obtained in numerical studies by \cite{diebels2002} for simple shear and \cite{tekoglu2011} for uniaxial compression. Studies which impose periodic boundary conditions at the sides (at $x=0, x=W$) show the same qualitative behaviour in compression, but a "smaller is stiffer" behaviour for simple shear (\cite{tekoglu2011} and \cite{tekoglu2005}). 
We observe this kind of response for systems with an aspect ratio of $H/W>2 R_0$ as seen in \figref{fig:aspect_ratio}. The explanation is that the clamped boundary locally increases the stiffness whereas the free surface decreases it (see Appendix B). With increasing system width the stiffening effect becomes more and more dominant and thus results in an "smaller is stiffer" size effect for wide and periodic systems.
 
\begin{figure}
\centering
\begin{minipage}[c]{1.0\textwidth}
\raisebox{-2.5em}{\rotatebox{90}{\small Compression}} 
\begin{minipage}[c]{.27\textwidth}
\centering
{\includegraphics[width=1.\textwidth]{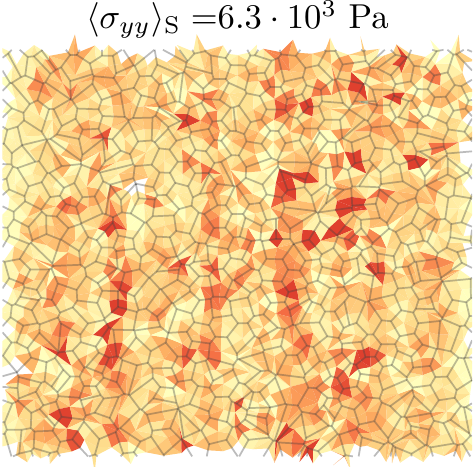}}
\end{minipage}
\hfill
\begin{minipage}[c]{0.27\textwidth}
\centering 
{\includegraphics[width=1.\textwidth]{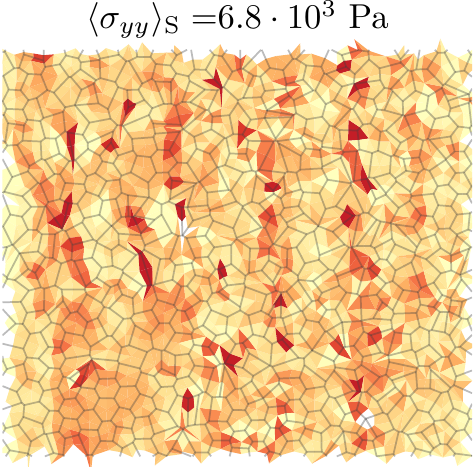}}
\end{minipage} 
\hfill
\begin{minipage}[c]{0.27\textwidth}
\centering 
{\includegraphics[width=1.\textwidth]{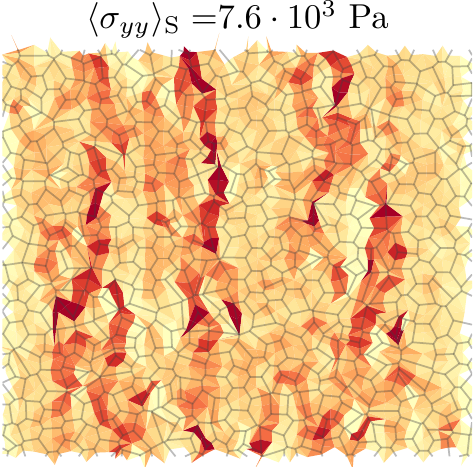}}
\end{minipage}
\begin{minipage}[c]{.14\textwidth}
\vspace{1em}
{\includegraphics[width=1.\textwidth]{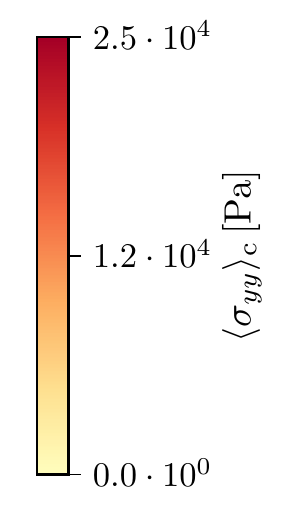}}
\end{minipage}
\end{minipage}
\begin{minipage}[c]{1.\textwidth}
\raisebox{-1.8em}{\rotatebox{90}{\small Simple shear}}
\begin{minipage}[c]{.27\textwidth}
\centering 
{\includegraphics[width=1.\textwidth]{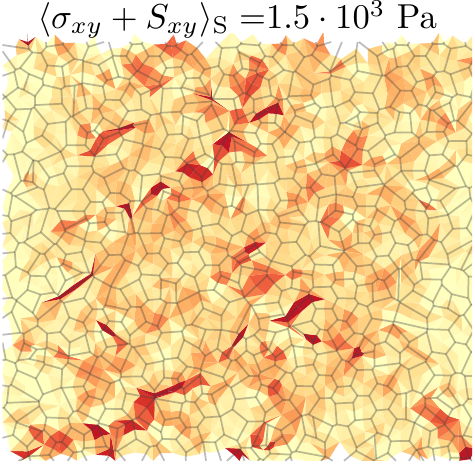}}
{\small Weakest systems}
\end{minipage}
\hfill
\begin{minipage}[c]{.27\textwidth}
\centering 
{\includegraphics[width=1.\textwidth]{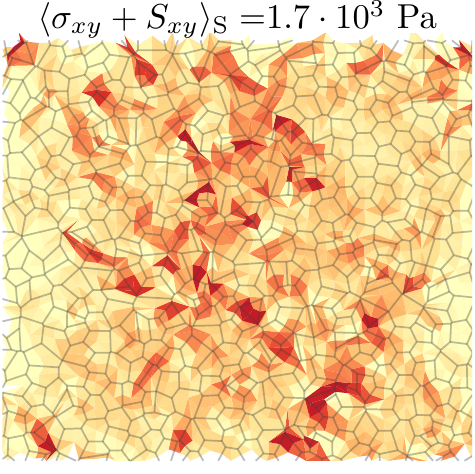}}
{\small Intermediate systems} 
\end{minipage}
\hfill
\begin{minipage}[c]{.27\textwidth}
\centering 
{\includegraphics[width=1.\textwidth]{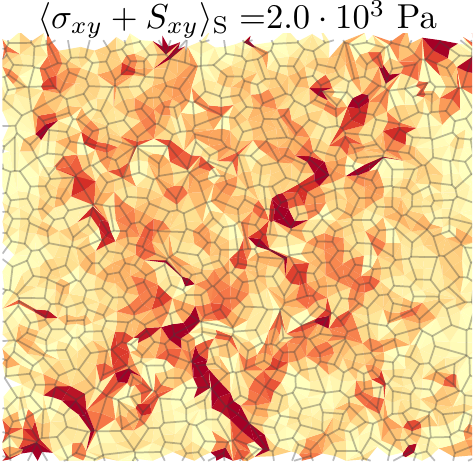}}
{\small Stiffest systems}
\end{minipage}
\begin{minipage}[c]{.14\textwidth}
\vspace{-0.8em}
{\includegraphics[width=1.0\textwidth]{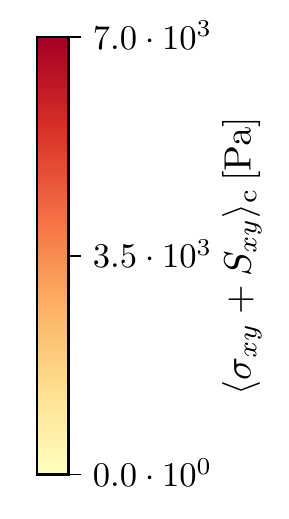}}
\end{minipage}
\end{minipage}
\caption{Local stress response for the weakest, stiffest and mean systems with $H=19 \Delta p$ and $\beta=0.3$. The top row shows the result for uniaxial compression, the bottom row for simple shear.\label{fig:stiff_stress}}\label{fig:micro_analysis}
\end{figure}

The system response of locally perturbed regular structures significantly differs depending on where the perturbations of the structure are located (cf. \figref{fig:sys_response_perturbated_microstructures}): for both loading cases a perturbation along the free boundaries of the system results in a macroscopically smaller stiffness compared to a perturbation in the center of the specimen or a perturbation following the constrained boundaries. This observation is also part of the explanation why, under compression, random systems are stiffer compared to regular systems. The surface weakening, which dominates in small samples, competes with bulk strengthening by a stiffer microstructure which results in the observed behavior: small irregular systems are weaker than regular ones, but large irregular systems are stiffer. 
To better understand the stiffening effect of disorder we study the stiffness variations observed between different specific realizations of a given random microstructures (cf. \figref{fig:subrealization}). Thus we ask: Among different microstructures constructed according to the same statistical rule, what makes some stiffer and others weaker? In compression, we find that stiffness differences correlate with a patterning of the stress distribution as shown in  \figref{fig:micro_analysis} for $H=19 \Delta p$ and $\beta=0.3$.  It can be seen that in the stiffest system the beams are oriented such that they form vertical columns which carry most of the load. The weak system shows fewer and less extended beam chains. 

\begin{figure}
	\centering
	\begin{minipage}[c]{0.88\textwidth}
	{\includegraphics[width=1.\textwidth]{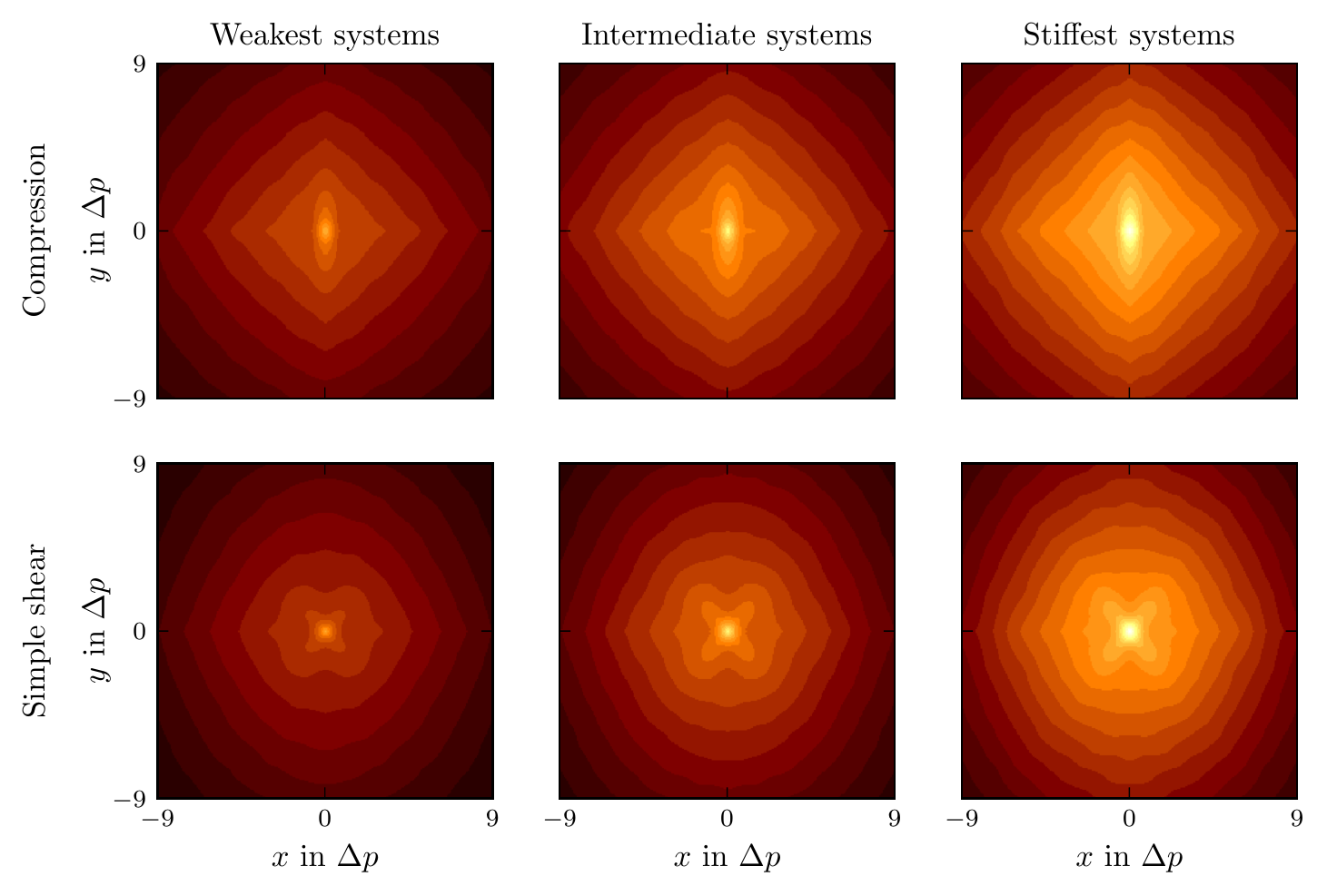}}
	\end{minipage}
	\begin{minipage}[c]{.11\textwidth}
		{\includegraphics[width=1.\textwidth]{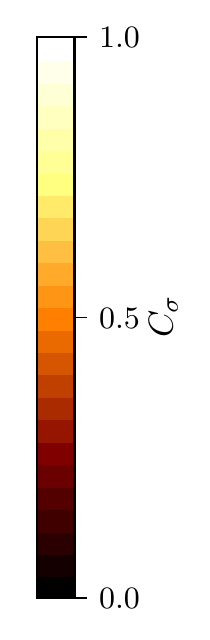}}
	\end{minipage}	
\caption{Stress auto-correlation functions for uniaxial compression and simple shear; averages over the ten weakest, ten intermediate and ten stiffest systems. System size $H=19 \Delta p$, disorder parameter $\beta=0.3$. For better comparison correlations are normalized by the maximum value of the stiffest system.}
\label{fig:correlation_analysis} 
\end{figure}

\begin{figure}
\centering
{\includegraphics{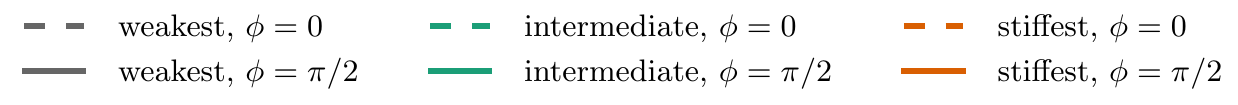}}
\begin{minipage}{1.\textwidth}
\hspace*{\fill}
{\includegraphics{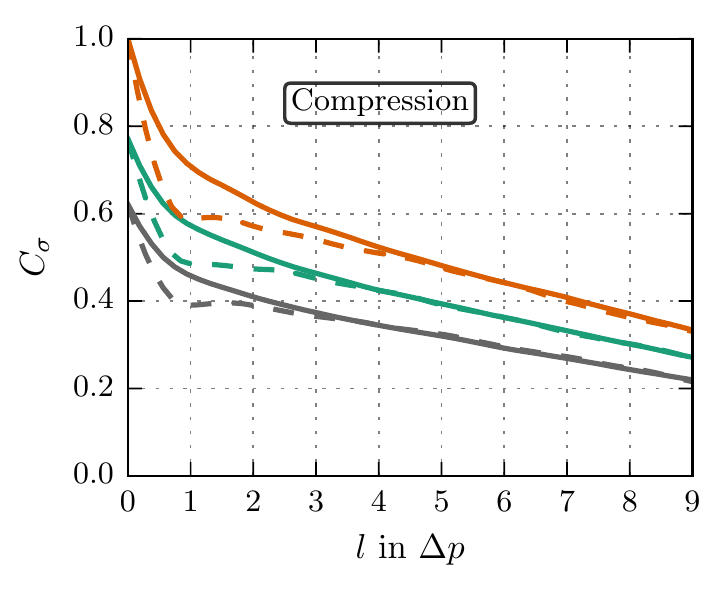}}
\hfill
{\includegraphics{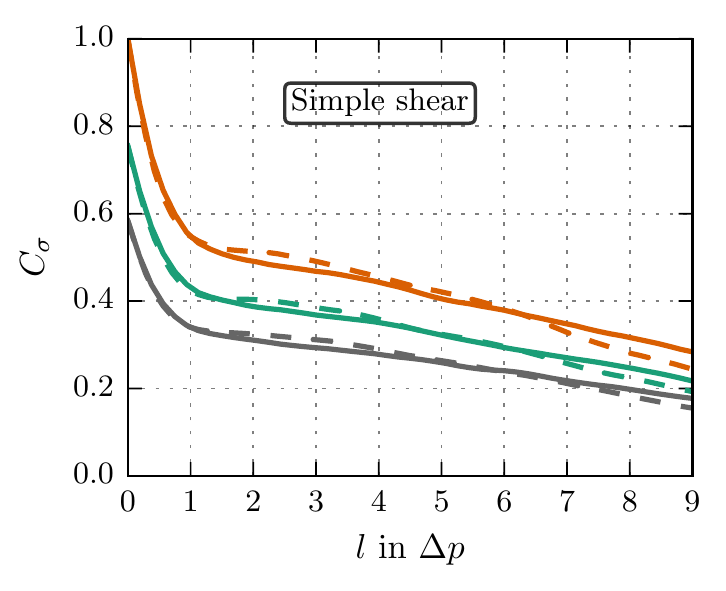}}
\hspace*{\fill}
\end{minipage}
\caption{Stress auto-correlations normalised with the maximum value of the stiffest systems along the vertical axis ($\phi=\pi/2$) and horizontal axis ($\phi=0$), for uniaxial compression and simple shear; averages over the ten weakest, ten intermediate and ten stiffest systems. System size $H=19 \Delta p$, disorder parameter $\beta=0.3$.} \label{fig:correlation_comparison}
\end{figure}

In order to quantify this observation we study the spatial correlation structure of the stress field in terms of the dominant stress tensor component (axial stress for compression, shear stress for simple shear):
\begin{align}
\sigma= \begin{cases} 
 \sigma_{yy}  & \text{for compressive loading},\\
 \sigma_{xy} + S_{xy}  & \text{for shear loading}.
 \end{cases}
\end{align}
As a quantitative characteristic of the spatial correlation structure we use the stress auto-correlation
\begin{align}
C_{\sigma} (\tens{l}) = \left\langle \left( \sigma(\tens{r}) - \langle \sigma  \rangle_\mrm{S} \right) \left( \sigma (\tens{r}+\tens{l}) - \langle \sigma  \rangle_\mrm{S} \right) \right\rangle_\mrm{S}, 
\end{align}
where $\tens{l} = [l \cos \phi, l \sin \phi]$ is the vector between two generic points $\tens{r}, \tens{r}+\tens{l}$, where $\phi$ is the angle with respect to the x-axis (\cite{Sandfeld2014_JStatMech}). To obtain good statistics, averages of ten simulations (the ten stiffest, ten weakest, and ten average microstructures) are used for the correlation analysis. For both loading cases it can be seen in \figref{fig:correlation_analysis} for $\beta=0.3$ and $H=19 \Delta p$ that stiffer systems show stronger and more extended stress auto-correlations. This allows us to quantify different microstructures based on their correlation functions to determine (macroscopically) stiffer or weaker systems. In addition it shows that for uniaxial compression the correlations are anisotropic and stronger in direction of the stress axis. This relates to the observation of force chains which, in analogy to granular systems, may be supposed to carry most of the load. Stiffer systems show stronger correlations, i.e. more pronounced force chains, and a less homogeneous stress distribution. By looking at the correlations along the vertical and horizontal axis (cf. \figref{fig:correlation_comparison}) one can see that the horizontal correlation has a dip at $l \approx \Delta p$. This means that the width of the force chains approximately corresponds to the width of one cell. 

Also for simple shear it can be observed that stiffer systems show a higher degree of correlation in their stress patterns. However, in comparison to compression the correlation is more isotropic. Furthermore the mechanisms governing local stiffness in this case are more complex and cannot simply be reduced to axial load transmission. Nevertheless a general conclusion can be drawn for both axial and shear deformation: More heterogeneous and more correlated stress patterns imply stiffer samples. 

\section{Conclusion}
\label{sec:conclusion}
With the increasing distribution of advanced processes, i.e., additive manufacturing, cellular structures become more and more important because of their superior weight specific properties. This demands a better understanding of the interplay between microstructural length scale, microstructure irregularity, and system size, and their effect on the macroscopic mechanical system response. These questions were studied in this work for randomly generated microstructures with different dimensions. We statistically analyzed large numbers of simulations, varying both morphological parameters (degree of randomness of the microstructure) and geometrical parameters (sample size and aspect ratio). It was found for simple shear and uniaxial compression that most investigated systems show a characteristic "smaller is weaker" behavior. However, the response under simple shear inverts to a "smaller is stiffer" behavior for wide systems as a result of the increasing influence of the boundary layer. 
This observation might be of interest for future (experimental) studies, as measured system responses cannot easily be interpreted as reflecting geometry independent material properties. From an engineering point of view this size dependent behavior is of great interest, especially for structures which dimensions are less than $\approx 50$ times the averaged cell size. It is important to note that this is not about the absolute size but rather about the ratio between system and cells, so that the studied size effects can appear in many natural or engineered structures.

To better understand deformation on the microstructure level, a continuization scheme was developed. The method provides a continuous representation of stresses and strains of beam networks and therefore allows to visualize local deformation patterns and to average them such as to allow for comparison with continuum models. It thus gives the possibility to statistically analyze and optimize microstructures. By analyzing both ordered and disordered microstructures in terms of the associated continuous representations of stress and strain, we could establish that stiffer microstructures exhibit both increased stress fluctuations and stronger spatial correlations of these fluctuations in the form of long stress transmission chains. These chains may be a quite general feature of disordered systems, see the recent work of \citet{Laubie2017_PRL} on disordered porous materials, and correlations in stress transmission chains may be relevant features in the run-up to fracture \citep{Laubie2017_PRL, Zaiser2017_P}. A deeper investigation of this issue that goes beyond linear elasticity may be an important topic of future studies. 

Beyond the scope of the present investigation, our continuization scheme may be a useful tool for further model development since it allows to directly compare results from discrete microstructure models with those deriving from continuum theories. This capability is of particular interest in view of higher order continuum theories which are also able to model size effects, but suffer from the problem that (higher order) constitutive parameters and boundary conditions are not straightforward to determine. Unlike most top-down attempts, where one takes the (size-dependent) macroscopic system response of the beam model and then tries to fit the parameters of the higher order continuum \citep[see e.g.][]{diebels2002, tekoglu2005, mora2007, Liebenstein2014_ProcApplMath} our approach allows to directly take the information of the microstructure into account. We showed this in an accompanying paper by \cite{liebenstein2017} where we identified constitutive parameters of a Cosserat continuum. 

\section*{Acknowledgements}
S.L. and M.Z. acknowledge funding by DFG under Grant no. 1 Za 171-9/1. M.Z. also acknowledges support by the Chinese government under the Program for the Introduction of Renowned Overseas Professors (MS2016XNJT044). 

\clearpage
\appendix
\section{Timoshenko Beam equations}
\label{appendix:eq_timoshenko_beam}
The governing equations of the Timoshenko beam can be found in the standard continuum mechanics literature, e.g. \cite{zienkiewicz2005b}.
The total displacements of the beam are a superposition of displacements $w_i$ along the beam axis (local coordinate $x_1$) and displacements caused by rotations $\phi_i$ of the beam cross-sections,
\begin{align}
\tens{u} = \begin{bmatrix}
w_1(x_1) - x_3 \phi_2(x_1)  +x_2 \phi_3(x_1) \\
w_2(x_1)  - x_3 \phi_1(x_1) \\
w_3(x_1)  + x_2 \phi_1(x_1)
\end{bmatrix}.
\end{align}
Note, that unlike in the Euler-Bernoulli beam theory $\phi_i \neq \tod{w_i}{x_1}$ and the rotation is considered an independent degree of freedom.
Disregarding body forces and dynamics the balance equations are given by
$\tod{F_i}{x_1} = 0$ ($i=1,2,3$) and 
\begin{align}
 &\dod{M_1}{x_1} = 0,&
 &\dod{M_2}{x_1} - F_3 = 0,&
 &\dod{M_3}{x_1} + F_2 = 0,
 \end{align} 
where $F_i$ are the beam force components and $M_i$ the bending moments acting around the axis $\tens{e}_i$. Besides the geometrical information regarding the cross-section area $A$ and the area moments of inertia 
\begin{align}
I_1 &= \int_A x_2^2 + x_3^2 \dif A,&
 I_2 &= \int_A x_3^2 \dif A, &
 I_3 &= \int_A x_2^2 \dif A,
 \end{align}
 we define the following constitutive parameters for an isotropic linearly-elastic beam: The Young's Modulus $E$, shear modulus $G$ and correction factors $k_i$ of the corresponding directions which depend on cross-section shape. The forces and moments can then be expressed in terms of the displacements as
 \begin{align}
 F_1 &= E A \dod{w_1}{x_1}, & F_2 &= k_2 G A \lb \dod{w_2}{x_1} - \phi_3\rb, & F_3 &= k_3 G A \lb \dod{w_3}{x_1}+\phi_2 \rb, \\
 M_1 &= k_1 G  I_1 \dod{\phi_1}{x_1}, & M_2 &= E I_2 \dod{\phi_2}{x_1}, &  M_3 &= E I_3 \dod{\phi_3}{x_1}.
 \end{align}
For a planar model as considered in the main paper these relations simplify since only the force and displacement components $F_{1,2}$ and $w_{1,2}$ need to be considered. The only rotation component that is relevant in the planar case is the angle $\phi_3 =: \phi$ and the associated moment $M_3 =: M$. Accordingly, the only shear correction factor that needs to be considered is $k_2 =: \kappa$. 

\section{Simple shear stress profiles}
\begin{figure}
\centering
\hfill
{\includegraphics{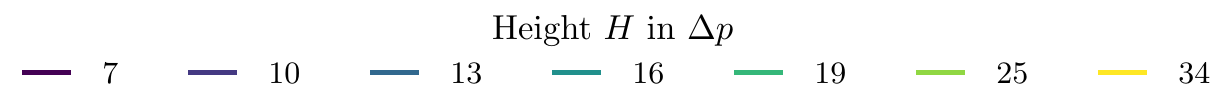}}
\hfill\vspace{-0.5em}
\hspace*{\fill}
\begin{subfigure}[c]{.49\textwidth}
{\includegraphics{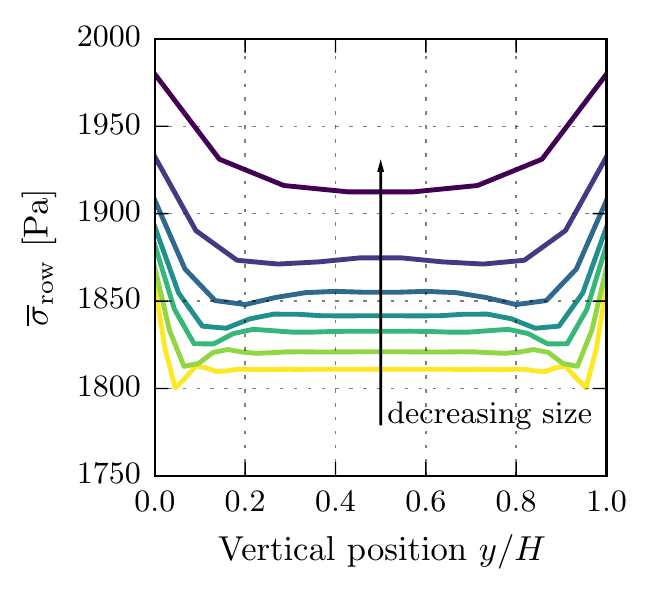}}
\caption{stress profiles along y-axis}\label{fig:stress_profiles_row_shear}
\end{subfigure}
\hfill
\begin{subfigure}[c]{.49\textwidth}
{\includegraphics{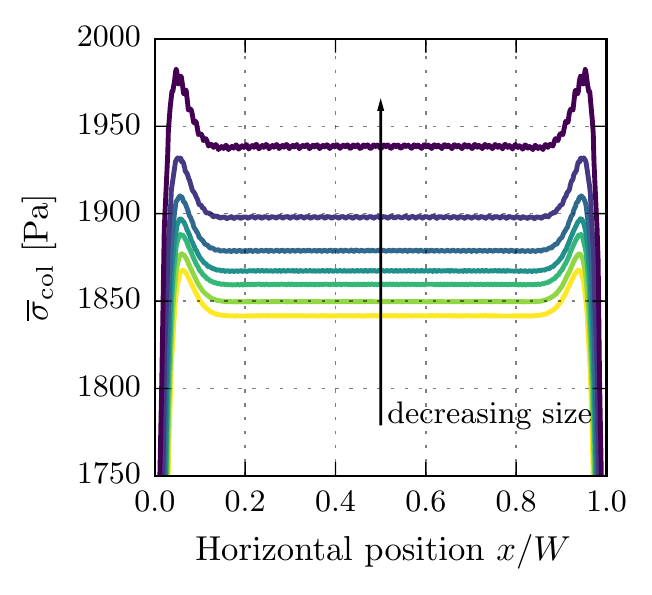}}
\caption{stress profiles along x-axis}
\label{fig:stress_profiles_col_shear}
\end{subfigure}
\hspace*{\fill}
\caption{Averaged stress profiles for simple shear of regular honeycomb structures of different sizes with aspect ratio $R=10 R_0$.}\label{fig:stress_profiles_shear}
\end{figure}

To better understand the size effect change for increasing aspect ratios under simple shear stress profile are analyzed. In analogy to the analysis for compression, the row and column-wise averages for simple shear loading are
\begin{align}
\overline{\sigma}_\mrm{row} &=  \frac{1}{N_\mrm{S} N_\mrm{CO} } \sum^{N_\mrm{S}} \sum^{N_\mrm{CO}} \frac{1}{W-2 a} \int_a^{W-a} \langle \sigma_{xy}+S_{xy} \rangle_\mrm{c}\dif x, \\
\overline{\sigma}_\mrm{col} &= \frac{1}{N_\mrm{S} N_\mrm{CO} }\sum^{N_\mrm{S}} \sum^{N_\mrm{CO}} \frac{1}{H-2 r} \int_r^{H-r} \langle \sigma_{xy}+S_{xy} \rangle_\mrm{c} \dif y.
\end{align} 
It can be seen in \figref{fig:stress_profiles_shear} that two competing effects exist. A stiffening at the constrained surfaces (top and bottom) and a softening at the free surfaces (left and right). Smaller system sizes show a significantly stiffer behavior at the constrained surfaces as well as in the bulk, which explains the observed "smaller is stiffer" behavior.

\section{Comparison of elastic strain energies}
\label{appendix:cauchy formulation}
\begin{figure}
\centering
\begin{minipage}[c]{0.88\textwidth}
\centering
{\rotatebox{90}{\small $\beta=0$}}
\begin{minipage}[c]{0.43\textwidth}
\centering
{\small Cauchy formulation}
{\includegraphics{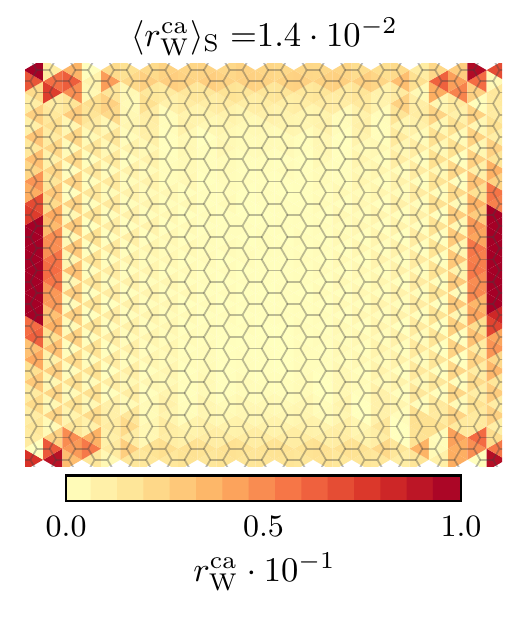}}
\end{minipage}
\begin{minipage}[c]{0.43\textwidth}
\centering
{\small Cosserat formulation}
{\includegraphics{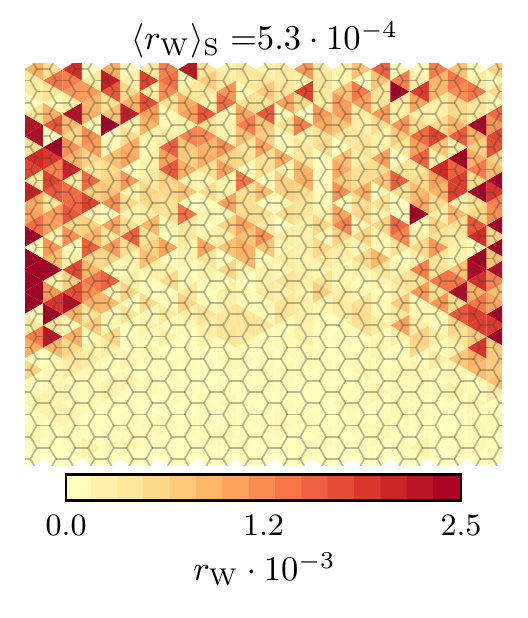}}
\end{minipage}
\end{minipage}
\begin{minipage}[c]{0.88\textwidth}
\centering
{\rotatebox{90}{\small $\beta=1$}}
\begin{minipage}[c]{0.43\textwidth}
\centering
{\includegraphics{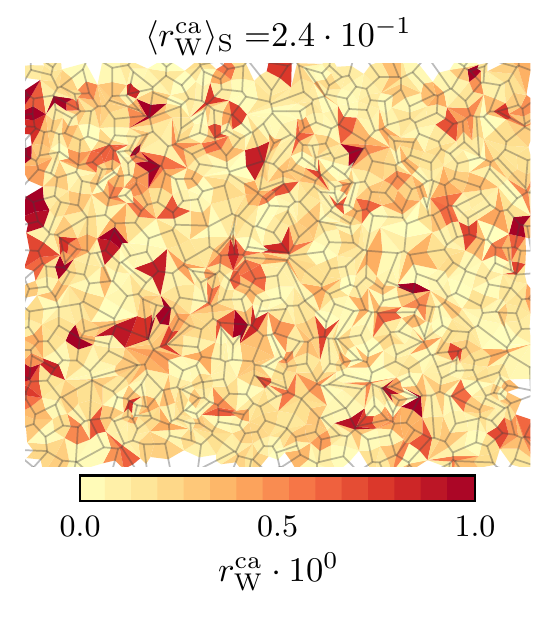}}
\end{minipage}
\begin{minipage}[c]{0.43\textwidth}
\centering
{\includegraphics{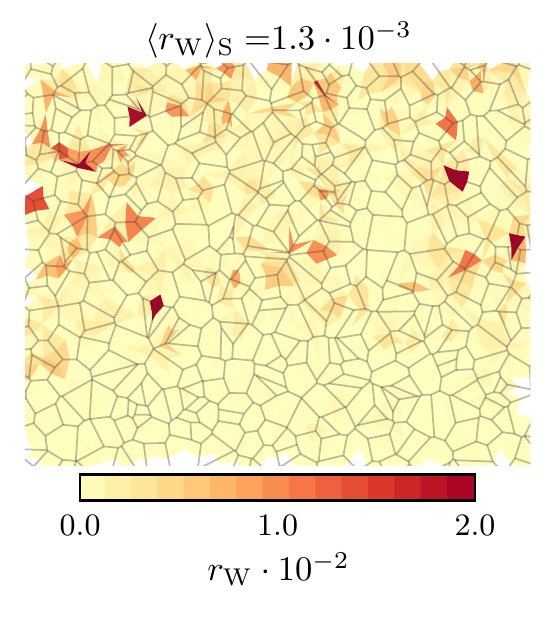}}
\end{minipage}
\end{minipage}
\caption{Comparison of energy residuals of classical Cauchy formulation (left) and Cosserat formulation (right) for regular (top) and random microstructures (bottom) under simple shear loading at system heights $H= 19 \Delta p$.  \label{fig:shear_energies}}
\end{figure}

Classical continuum formulations do not take the additional micro rotation into account. As a result they are not suitable for homogenization of discrete networks on the meso-scale (the scale of one control volume) but also on the macro-scale (the scale of the system). To illustrate this point we compare the classical Cauchy strain energy  
\begin{align}
W_\mrm{c}^\mrm{ca} = \frac{1}{2} \langle\tens{\sigma} \rangle_\mrm{c} :  \langle \tens{\varepsilon} \rangle_\mrm{c}
\end{align}
to our formulation \eqref{eq:strain_energy}.
Again as a measure a relative energy residual 
\begin{align}
r_\mrm{W}^\mrm{ca} &= \frac{| W_\mrm{c}^\mrm{ca}-W_\mrm{b,c} | }{W_\mrm{b,c}}
\end{align}
is chosen. In \figref{fig:shear_energies} the two energy residuals $r_\mrm{W}^\mrm{ca},r_\mrm{W}$ are compared for a regular and a irregular system under simple shear loading. For the simple shear case it can be seen that the maximum error for the irregular system in the Cosserat formulation, is in the range of $2\%$, whereas the overall error, similar to the case of compressive loading, is about $0.1\%$. In the Cauchy formulation which neglects rotations, by contrast, maximum errors are of the order of 1 whereas the average error is of the order of 0.25, rendering such formulations practically useless. As the energy residuals obtained from the Cosserat formulation are about two orders of magnitude smaller than those obtained from the Cauchy formulation we conclude that, for the structures at hand, Cosserat models out-perform classical Cauchy formulations of stress and strain by a very considerable margin.

\clearpage
\bibliographystyle{elsarticle-harv} 
\bibliography{bibfile}

\end{document}